\documentclass[11pt, a4paper]{article}
\usepackage{jcappub}
\usepackage{here}
\usepackage{graphicx}
\usepackage{amsmath,amsthm,amssymb}
\usepackage{mathrsfs,mathtools}
\usepackage{bm}
\usepackage{xcolor}
\usepackage[normalem]{ulem}
\usepackage{physics}
\usepackage{comment}
\usepackage{acronym}
\usepackage{siunitx}
\usepackage{xspace}
\usepackage{booktabs}

\usepackage{amsfonts}
\usepackage{dcolumn}

\hypersetup{colorlinks=true
,urlcolor=DARKBLUE
,anchorcolor=DARKBLUE
,citecolor=DARKBLUE
,filecolor=DARKBLUE
,linkcolor=DARKBLUE
,menucolor=DARKBLUE
,linktocpage=true
,pdfproducer=medialab
,pdfa=true
}
\allowdisplaybreaks[1]

\usepackage{stackengine}

\acrodef{PDF}{probability 
density function}
\acrodef{CMB}{cosmic microwave background}
\acrodef{PBH}{primordial black hole}
\acrodef{USR}{ultra-slow-roll}
\acrodef{MS}{Mukhanov--Sasaki}

\definecolor{MONZA}{HTML}{CF000F}
\definecolor{DARKBLUE}{HTML}{00008b}
\definecolor{DARKMAGENTA}{HTML}{8b008b}

\newcommand{\calC}{\mathcal{C}}
\newcommand{\ue}{\mathrm{e}}
\newcommand{\calO}{\mathcal{O}}
\newcommand{\calP}{\mathcal{P}}
\newcommand{\calR}{\mathcal{R}}
\newcommand{\us}{\mathrm{s}}

\newcommand{\Mpl}{M_\mathrm{Pl}}
\newcommand{\dip}{\mathrm{dip}}
\newcommand{\SR}{\mathrm{SR}}
\newcommand{\peak}{\mathrm{peak}}

\newcommand{\efolding}{$e$-folding\xspace}
\newcommand{\efolds}{$e$-folds\xspace}

\newcommand{\beae}[1]{\begin{equation}\begin{aligned} #1 \end{aligned}\end{equation}}
\newcommand{\bege}[1]{\begin{equation}\begin{gathered} #1 \end{gathered}\end{equation}}
\newcommand{\bae}[1]{\begin{align} #1 \end{align}}

\newcommand{\bmte}[1]{\begin{multlined}[t] #1 \end{multlined}}

\title{Dip and non-linearity in the curvature perturbation from inflation with a transient non-slow-roll stage}

\author[a,b]{Tomohiro Fujita,}
\author[c]{Ryodai Kawaguchi,} 
\author[b,d,e,f]{Misao Sasaki,}
\author[g,h]{and Yuichiro Tada}

\affiliation[a]{Department of Physics, Ochanomizu University, Bunkyo, Tokyo 112-8610, Japan}
\affiliation[b]{Kavli Institute for the Physics and Mathematics of the Universe (WPI),
The University of Tokyo Institutes for Advanced Study,
The University of Tokyo, Chiba 277-8583, Japan}
\affiliation[c]{Department of Physics, Waseda University, 3-4-1 Okubo, Shinjuku, Tokyo 169-8555, Japan} 
\affiliation[d]{Asia Pacific Center for Theoretical Physics, Pohang 37673, Korea} 
\affiliation[e]{Center for Gravitational Physics and Quantum Information,
Yukawa Institute for Theoretical Physics, Kyoto University, Kyoto 606-8502, Japan}
\affiliation[f]{Leung Center for Cosmology and Particle Astrophysics,
National Taiwan University, Taipei 10617, Taiwan}
\affiliation[g]{Institute for Advanced Research, Nagoya University,
Furo-cho Chikusa-ku, 
Nagoya 464-8601, Japan}
\affiliation[h]{Department of Physics, Nagoya University, 
Furo-cho Chikusa-ku,
Nagoya 464-8602, Japan}

\emailAdd{fujita.tomohiro@ocha.ac.jp}
\emailAdd{ryodai0602@fuji.waseda.jp}
\emailAdd{misao.sasaki@ipmu.jp}
\emailAdd{yuichiro.tada@rikkyo.ac.jp}

\abstract{
We consider models of inflation that contain a transient non-slow-roll stage and investigate the conditions under which a dip appears in the power spectrum of the curvature perturbation.
Using the $\delta N$ formalism, 
we derive a general relation between the comoving curvature perturbation $\calR$ and the scalar field perturbation $\delta\varphi$ and its velocity perturbation $\delta\pi$.
Compared with the result obtained in linear perturbation theory,
it turns out that properly taking account of the $\delta\pi$ contribution is essential to reproduce the dip in the power spectrum. Namely, the curvature perturbation is proportional to a specific linear combination of $\delta\varphi$ and $\delta\pi$ at the linear order.
We also investigate the non-linearity at the dip scale and find that models with a bump or an upward step exhibit much larger non-linearity than ultra-slow-roll and Starobinsky's linear potential models.
Finally, we demonstrate the importance of non-linearity by computing the probability 
density functions (PDFs) for the models mentioned above and show that highly asymmetric PDFs are realised for models with a bump or a step.
}

\begin{document}

\newcommand{\newc}{\newcommand}
\newcommand{\rk}{\textcolor{red}}

\newc{\be}{\begin{equation}}
\newc{\ee}{\end{equation}}
\newc{\D}{\partial}
\newc{\rH}{{\rm H}}
\newc{\rd}{{\rm d}}
\newcommand{\rBH}{r_{s}}
\newcommand{\rc}{r_{c}}
\newcommand{\rh}{r_{h}}
\newcommand{\Xh}{X_{h}}
\newcommand{\hr}{\hat{r}}
\newcommand{\ma}[1]{\textcolor{magenta}{#1}}
\newcommand{\cy}[1]{\textcolor{cyan}{#1}}
\renewcommand{\mm}[1]{\textcolor{red}{#1}}

\begin{flushright}
WUCG-25-03 YITP-25-21
 \\
\end{flushright}

\maketitle
\acresetall

\section{Introduction}
\label{Introduction}
The theory of inflation 
describes the universe at its very early stage that undergoes an era of exponentially accelerated expansion~\cite{Brout:1977ix,Starobinsky:1980te,Sato:1980yn,Kazanas:1980tx,Guth:1980zm,Linde:1981mu,Albrecht:1982wi}.
Its most important prediction is that, by stretching quantum vacuum fluctuations to super Hubble scales, it explains the origin of all the inhomogeneous structures of the universe today~\cite{Mukhanov:1981xt,Bardeen:1983qw,Mukhanov:1985rz,Sasaki:1986hm}.

Numerous inflation models have been proposed so far.
One of their standard models is the slow-roll 
one with a single scalar field called inflaton (see, e.g., Refs.~\cite{Starobinsky:1980te,Linde:1983gd,Freese:1990rb,Bezrukov:2007ep,Kallosh:2013yoa}).
In these models, called slow-roll inflation models, the inflaton has a potential with a mild slope and rolls slowly over it.
The inflationary universe is realized as the potential energy of the inflaton effectively plays the role of a cosmological constant.
However, we have not yet identified the inflaton,  
which remains a major challenge for modern cosmology.

In order to identify the inflaton, it is necessary to understand the observational signals originating from inflation.
The best understanding of inflation so far has been provided by observations of the \ac{CMB} anisotropies by the Planck collaboration~\cite{Planck:2018vyg,Planck:2018jri}.
Their results imply that the curvature perturbation on comoving slices $\calR$ is highly Gaussian and its spectrum is almost scale-invariant on large scales ($\sim\SI{10}{Mpc}\text{--}\SI{1}{Gpc}$),
which is perfectly consistent with the theoretical predictions of slow-roll inflation models.

On the other hand, 
observational understanding on small scales is limited.
It has not been confirmed if smaller-scale perturbations are also almost Gaussian and scale-invariant or not.
There is no doubt that information on small scales will play an important role in unravelling the full story of inflation and in identifying the high-energy physics behind it.
On the theoretical side, it is important to investigate what kind of signals inflation can leave on small scales. 

In this paper, we focus on very small scales ($\ll \SI{1}{Mpc}$).
Specifically, we consider models of inflation that have a transient non-slow-roll stage and examine the signals from such models.
During the non-slow-roll stage, the inflaton velocity deviates substantially from that given by the slow-roll condition.
Such models have been studied extensively in the context of \ac{PBH} ~\cite{zel1967hypothesis,Hawking:1971ei,Carr:1974nx,Carr:1975qj} production, because 
they can amplify the power spectra of the curvature perturbation on small scales
(see Refs.~\cite{Sasaki:2018dmp,Carr:2020xqk,Green:2020jor,Carr:2020gox,Villanueva-Domingo:2021spv,
Carr:2021bzv,Escriva:2022duf,Karam:2022nym,Ozsoy:2023ryl} for recent reviews on \acp{PBH}).

In this paper, we do not focus on the scale where the power spectrum is amplified, i.e., the peak scale, but rather on the scale at which the power spectrum has a minimum, a dip.
In those models where the power spectrum is enhanced by a factor of $\sim10^{8\text{--}9}$, the dip is at the scale $\sim10^{2\text{--}3}$ times larger than the peak scale. 
It is a feature at the largest scale on which the non-slow-roll stage can leave its effect. 
The existence of a dip has been noted in several studies (see, e.g., Refs.~\cite{Leach:2001zf,Byrnes:2018txb,Tasinato:2020vdk,Cheng:2023ikq}), and its properties have been discussed recently in the context of linear perturbation theory~\cite{Wang:2024wxq,Briaud:2025hra}.
In this study, using the $\delta N$ formalism~\cite{Salopek:1990jq,Sasaki:1995aw,Starobinsky:1985ibc,Sasaki:1998ug,Lyth:2004gb,Lee:2005bb,Lyth:2005fi,Abolhasani:2019cqw}, we investigate the dip in a more general manner and clarify the condition for the existence of a dip at the non-linear level in the transient non-slow-roll models. 
We also suggest that the dip may disappear due to non-linear effects in some of these models.

We discuss the non-Gaussianity of the curvature perturbation at the dip scale and compute its \ac{PDF} as well, taking advantage of the $\delta N$ formalism, a non-perturbative analysis.
Non-Gaussianity in the context of inflation has been investigated for a long time~\cite{Komatsu:2001rj,Bartolo:2001cw,Maldacena:2002vr,Acquaviva:2002ud,Bartolo:2004if,Chen:2010xka} and has been widely studied in models that violate the slow-roll condition (see, e.g., Refs.~\cite{Namjoo:2012aa,Chen:2013aj,Chen:2013eea,Martin:2012pe,Cai:2018dkf}).
In particular, there has been a lot of research in recent years on calculating the PDF of the curvature perturbation and their impact on the PBH abundance~\cite{Atal:2019cdz,Atal:2019erb,Biagetti:2021eep,Kitajima:2021fpq,Gow:2022jfb,Ferrante:2022mui,Pi:2022ysn} (see also Refs.~\cite{Pattison:2017mbe,Ezquiaga:2019ftu,Figueroa:2020jkf,Pattison:2021oen,Ahmadi:2022lsm,Animali:2022otk,Vennin:2024yzl} for the so-called stochastic inflation approach to these problems).

Some of us calculated \ac{PDF} at the dip scale in the upward step model in a previous study and showed the existence of a highly asymmetric PDF that largely deviates from the Gaussian distribution~\cite{Kawaguchi:2023mgk}.
The highly asymmetric PDF found there shows a distribution biased towards positive values of $\calR$ while it almost vanishes in the negative value region.
In this study, we extend our previous work in two directions.
First, we incorporate the contribution not only from $\delta\varphi$ but also from its velocity perturbation $\delta\pi$.
It turns out that the $\delta\pi$ contribution is crucial at the dip scale. 
Hence, for the accurate calculation of the PDF, we need to incorporate the $\delta\pi$ contribution.
Second, we perform analyses of several different types of non-slow-roll stage models and compare their results.
We find that, in the \ac{USR} and Starobinsky's linear potential models, the $\delta\pi$-originated linear term is the main contribution at the dip scale, and their curvature perturbations remain mostly Gaussian.
In contrast, in the bump and step models, the non-linearity makes a significant contribution and the \ac{PDF} is highly asymmetric, similar to the result of \cite{Kawaguchi:2023mgk}, but a non-negligible probability for negative values of $\calR$ appears due to the $\delta\pi$ contribution.

This paper is organised as follows.
In Sec.~\ref{model}, we introduce a solvable inflationary model with a canonical scalar field. 
In Sec.~\ref{Appendixpowerspectrum}, we derive the linear perturbation power spectrum for the three-stage model (i.e., a slow-roll stage followed by a transient non-slow-roll stage, followed by another slow-roll stage) and visualise it for several sets of model parameters.
In Sec.~\ref{PSofdeltaN}, we obtain the analytical expression for the curvature perturbation by using the $\delta N$ formalism. 
This procedure suggests that the time fluctuation $\delta N$ is most conveniently expressed in terms of a new set of variables $\delta X$ and $\delta Y$ that consist of linear combinations of $\delta\varphi$ and $\delta\pi$.
In Sec.~\ref{linearanalysis}, we compare the power spectrum of the curvature perturbation from the $\delta N$ formalism with that from the linear perturbation theory and reveal a condition under which a dip in the power spectrum appears.
In Sec.~\ref{nonlinearity}, we illustrate the \ac{PDF} of the curvature perturbation at the dip scale and discuss the importance of the non-perturbative calculation for the four different models.
A parameter characterising the non-linearity of the curvature perturbation is also defined, giving us an intuitive understanding of the condition under which the curvature perturbation becomes highly non-Gaussian.
Sec.~\ref{conclusion} is devoted to conclusions. 

\section{Solvable three-stage model}
\label{model}
In this section, we describe the setup of our model.
We consider a simple toy model in which the equations of motion can be solved analytically and which can incorporate various types of transient non-slow-roll stages.
Let us start with the action of a canonical single scalar field:
\be
\mathcal{S}=\int \dd[4]{x} \sqrt{-g} \left[\frac{\Mpl^{2}}{2}R-\frac{1}{2}g^{\mu\nu}\partial_{\mu}\phi\partial_{\nu}\phi-V(\phi)\right]\,,
\label{action}
\ee
where $g$ is a determinant of the metric tensor $g_{\mu\nu}$, $R$ is the corresponding Ricci scalar, and $V(\phi)$ is a potential of the scalar field.
We consider a spatially flat Friedmann--Lema\^{i}tre--Robertson--Walker background, $\dd{s}^2=-\dd{t}^2+a^2(t)\delta_{ij}\dd{x^i}\dd{x^j}$, where $a(t)$ is a time-dependent scale factor.
From the action~\eqref{action}, we obtain the Friedmann equation and equation of motion for the scalar field, 
\bae{3h^{2}=\frac{1}{2}h^{2}\pi^{2}+v\,\qc
\dv{\pi}{n}+\frac{v}{h^{2}}\pi+\frac{\partial_\varphi v}{h^{2}}=0\,,
}
where we introduce dimensionless quantities, $\varphi\equiv\phi/\Mpl$, $\pi\equiv \dv*{\varphi}{n}$, $v\equiv V(\phi)/V_0$, and $h\equiv \Mpl H/\sqrt{V_0}$.
$V_0$ is an arbitrary reference point of the potential, $H=\dot{a}/a$ is the Hubble expansion rate and $n$ is the number of \efolds defined by $\dd{n}=H\dd{t}$.
Combining those two equations and assuming the first slow-roll condition $\pi^2\ll6$, the equation of motion reduces to
\be
\dv{\pi}{n}+3\pi+3\frac{\partial_
{\varphi} v}{v}=0.
\label{backgroundeom}
\ee
Without loss of generality, we can assume that $\varphi$ moves from positive to negative along the potential and $\pi$ is always negative during inflation. 
We use this assumption in the following part of the paper.

In order to incorporate various single-field inflation models, we divide the inflaton potential into three stages\footnote{A similar analysis with a two-stage version has been performed in Refs.~\cite{Domenech:2023dxx,Wang:2024wxq}.
In this paper, we extend the model to three stages in order to include, for example, the bump/step models.} and introduce constants $A_i$ in each stage as 
\be
\frac{\partial_\varphi v}{v}=
\begin{cases}
A_1  
& (\varphi>\varphi_+)\,, \\ 
A_2 
& (\varphi_-<\varphi\le\varphi_+)\,,
\\
A_3 
& (\varphi\le\varphi_-)\,,
\end{cases}
\label{potential}
\ee
where $\varphi_+$ and $\varphi_-$ are scalar field values at boundaries between the first and second stages, and the second and third stages, respectively.
The model potential is exponential~\cite{Lucchin:1984yf,Andrianov:2011fg}, where $A_i$ corresponds to the index.
In the slow-roll context, $\abs{A_i}$ is related to the potential-type slow-roll parameter as $\abs{A_i}=\abs{(\partial_\varphi v)/v}\eqqcolon\sqrt{2\epsilon_V}$.
In this paper, we assume that $A_1$ and $A_3$ are positive but small enough ($\ll1$) to realise standard slow-roll phases.
On the other hand, we allow $A_2$ to have any values, so that our setup includes various types of inflaton potential, e.g, \ac{USR} potential (see, e.g., Refs.~\cite{Garcia-Bellido:2017mdw,Kannike:2017bxn,Germani:2017bcs,Ezquiaga:2017fvi,Motohashi:2017kbs}), Starobinsky linear potential~\cite{Starobinsky:1992ts,Ivanov:1994pa,Pi:2022zxs}, bump/step potential (see, e.g., Ref.~\cite{Inomata:2021tpx,Cai:2021zsp,Cai:2022erk,Wang:2024nmd}), etc., violating the second slow-roll condition (i.e., the time variation of the first slow-roll parameter is large).

\begin{table} 
    \renewcommand{\arraystretch}{1.3}
    \centering
    \begin{tabular}{cccccc} 
        \toprule
        Model & $A_1$ & $A_2$ & $A_3$ & $N_2$ & $\pi_-$ \\ \hline 
        USR & $0.01$ & $0$ & $0.01$ & $2.57$ & $-4.48\times10^{-6}$ \\
        Starobinsky linear & $0.01$ & - & $\num{4e-6}$ & - & $-0.01$ \\
        Bump & $0.01$ & $-0.015$ & $0.01$ & $1.7019\times10^{-1}$ & $-3.83\times10^{-6}$ \\
        Step & $0.01$ & $-120$ & $0.01$ & $2.77695\times10^{-5}$ & $-2.56\times10^{-6}$ \\
        \bottomrule
    \end{tabular}
    \caption{Four sets of model parameters. $A_i$ corresponds to the index of the exponential potential for each stage. $\pi_-$ is the velocity at $\varphi=\varphi_-$. $N_2$ denotes the time duration of the second piece of the potential}
    \label{table1}
\end{table}
\begin{figure} 
    \centering
        \includegraphics[width=0.9\hsize]{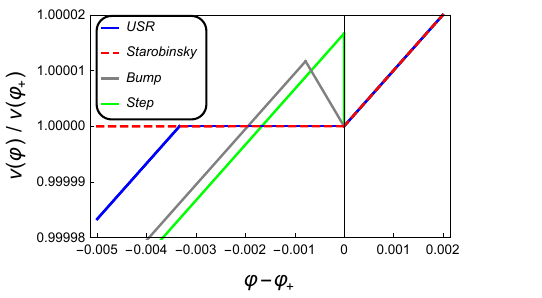}   
    \caption{\label{Fig_potential} 
    The shape of the inflaton potential $v(\varphi)$ for the four different models indicated in the legend, with the parameter sets listed in Table~\ref{table1}. All the models have the same potential at $\varphi>\varphi_+$.
    We adopt those parameter sets throughout the paper.
    The vertical axis is normalised by $v(\varphi_+)$.
    }
\end{figure}
In the following sections, we first derive the analytic formula of $\delta N$ for the general three-stage potential~\eqref{potential} and then investigate the \ac{PDF} of curvature perturbation around the dip scale.
Specifically, we will focus on the USR potential, Starobinsky's linear potential, bump potential and step potential (see Fig.~\ref{Fig_potential} for the potential shapes).

\section{Power spectrum in linear perturbation theory}
\label{Appendixpowerspectrum}
Before proceeding with the analysis using the $\delta N$ formalism, let us calculate the power spectrum of the curvature perturbation in linear perturbation theory. 
In discussing the inflaton perturbation, the \ac{MS} variable $u$ is commonly used.
It relates to the comoving curvature perturbation ${\calR}$ as $u=a\Mpl\pi{\calR}$ and to the scalar field perturbation $\delta\varphi$ in the spatially flat slicing as $u=-a\Mpl\delta\varphi$.
The mode function of MS variable $u_k$ follows the MS equation~\cite{Bardeen:1980kt,Kodama:1984ziu,Mukhanov:1985rz,Sasaki:1986hm},
\be
u_k''+\left(k^2-\frac{Z''}{Z}\right)u_k=0,
\label{MSeq}
\ee
where $Z=a\Mpl\pi$.
In our setup~\eqref{potential}, $Z''/Z$ can be recast as
\be
\frac{Z''}{Z}=\frac{2}{\tau^2}+{\cal{O}}(\pi^2),
\ee
with the conformal time $\tau$, except at the transition points.
Here, we use the approximation that $\pi^2$ is much smaller than unity and $\tau\simeq-1/(aH)$.
The mode function can then be written by elementary functions as
\be
u_k=i\sqrt{\frac{1}{2k}}\frac{1}{k\tau}\left[\alpha^{(i)}_k(1+ik\tau)e^{-ik\tau}+\beta^{(i)}_k(1-ik\tau)e^{+ik\tau}\right]\,,
\ee
where $\alpha_k^{(i)}$ and $\beta_k^{(i)}$ are constants in each stage ($i=1,2,3$).
By imposing the adiabatic vacuum initial condition \cite{Birrell:1982ix} and appropriate boundary conditions at $\varphi_+$ and $\varphi_-$, 
the constants are given by 
\begin{itemize}
\item (1st stage)
\be
\alpha_k^{(1)}=-1\, 
\qc
\beta_k^{(1)}=0 \,, 
\ee
\item (2nd stage)
\be
\alpha_k^{(2)}=-1+\frac{3i(A_1-A_2)k_{+}(k^2+k_+^2)}{2A_1 k^3}\, \qc 
\beta_k^{(2)}=\frac{3i(A_1-A_2)e^{2i\frac{k}{k_+}}(k+ik_+)^2k_+}{2A_1 k^3}\,, 
\ee
\item (3rd stage)
\bae{
&\begin{aligned}
\alpha_k^{(3)}=\bmte{\frac{1}{4A_1 k^6 \pi_-}
\biggl[9(A_1-A_2)(A_2-A_3)e^{-2ik\left(\frac{1}{k_-}-\frac{1}{k_+}\right)}(k-ik_-)^2k_- (k+ik_+)^2 k_+
\\
-\left(2A_1 k^3-3i(A_1-A_2)k_+(k^2+k_+^2)\right)\left(3i(A_2-A_3)k_-(k^2+k_-^2)+2k^3\pi_-\right)
\biggr]
\,,}
\end{aligned}
\\
&\begin{aligned}
\beta_k^{(3)}=\bmte{\frac{1}{4A_1 k^6 \pi_-}
\biggl[3(A_2-A_3)e^{2i\frac{k}{k_-}}(k+ik_-)^2k_-\left(-3(A_1-A_2)(k^2+k_+^2)k_+\right)
\\
+3(A_1-A_2)e^{2i\frac{k}{k_-}}(k+ik_+)^2k_+\left(3(A_2-A_3)k_-(k^2+k_-^2)+2ik^3\pi_-\right)\biggr]
\,,}
\end{aligned}
}
\end{itemize}
where $\pi_\pm$ and $k_\pm$ denote the velocities and scales exiting the Hubble horizon when $\varphi=\varphi_\pm$ respectively.

\begin{figure} 
    \centering
    \includegraphics[width=0.9\hsize]{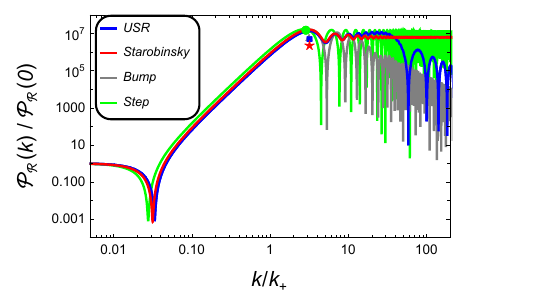}
    \caption{\label{powerspectrum} 
    The power spectrum of the curvature perturbation ${\calR}$ for the four different models with model parameters shown in Table~\ref{table1}.
    The vertical axis is normalised by the power spectrum at $k\to0$ and the horizontal axis is normalised by $k_+$.
    The blue square, red star, gray triangle, and green circle denote the peak values of the corresponding models calculated by the approximate formula~\eqref{approximated_power} for the power spectrum.}
\end{figure}

The (dimensionless) power spectrum of curvature perturbation at $\tau\to0$ is given by
\be
{\cal{P}}_{\calR}(k)=\frac{H^2}{4\Pi^2 \Mpl^2 A_3^2}|\alpha_k^{(3)}+\beta_k^{(3)}|^2\,.
\label{Apppowerspectrum}
\ee
Here, to distinguish it from the scalar field velocity $\pi$, the Pi ($\approx 3.14$) is denoted by $\Pi$. 
We show the power spectrum~\eqref{Apppowerspectrum} of curvature perturbation calculated by the linear perturbation theory in Fig.~\ref{powerspectrum}.
We use four concrete models, USR (blue line), Starobinsky linear  (red line), bump (gray line), and step (green line) models, of which parameters and plots are shown in Table~\ref{table1} and Fig.~\ref{Fig_potential}.
Here, the duration of the non-slow-roll stage, $N_2$, is treated as a free parameter instead of $\varphi_+-\varphi_-$.
For given $A_i$'s and $N_2$, $\pi_-$ can be determined in a manner consistent with the background equation of motion~\eqref{backgroundeom}, and the results are also shown in Table~\ref{table1}.
In each model, the parameters are adjusted so that the peak value of the power spectrum is about $10^7$ times larger than the value in the long wavelength limit $\calP_\calR(k\to0)$.
It is worth mentioning that the power spectrum in the long wavelength limit and that at the peak scale are approximately given by
\be
\calP^{(\SR)}_{\calR}\equiv\frac{H^2}{4\Pi^2 \Mpl^2 \pi_+^2}\, 
\qc
\calP^{(\peak)}_{\calR}\equiv\left(\frac{\bar{A}\pi_+}{A_3\pi_-}\right)^2\calP^{(\SR)}_{\calR}\,,
\label{approximated_power}
\ee
where $\bar{A}$ is the weighted average of $A_1$ and $A_3$,
\be
\bar{A}\equiv A_1 e^{-3N_2}+A_3(1-e^{-3N_2})\,.
\label{weighted_average}
\ee
In Fig.~\ref{powerspectrum}, we also plot the peak values evaluated by Eq.~\eqref{approximated_power}.
As we can see, they roughly agree with the results of linear perturbation theory.

\section{\boldmath Perturbative expansion of $\delta N$}
\label{PSofdeltaN}
In this section, we first derive the analytical solution on superhorizon scales in our setup to apply the $\delta N$ formalism. 
Then, to understand what happens around the dip, we compare the power spectrum calculated by the $\delta N$ formalism with the result from linear perturbation theory.

\begin{figure} 
    \centering
    \begin{tabular}{c}
        \begin{minipage}{0.95\hsize}
            \centering
            \includegraphics[width=
            0.7\hsize]{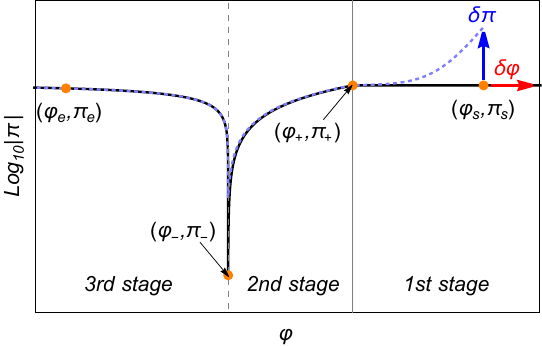}
        \end{minipage}
        \\
        \\
        \begin{minipage}{0.95\hsize}
            \centering
            \includegraphics[width=
            0.7\hsize]{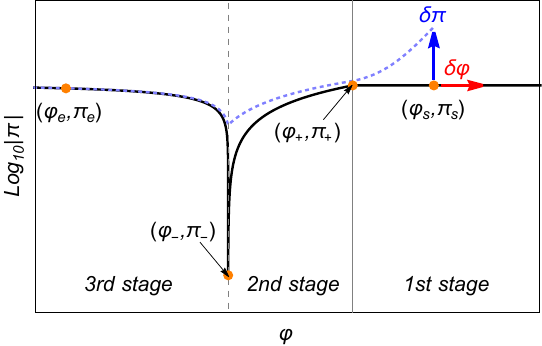}
        \end{minipage}
    \end{tabular}
    \caption{\label{schematic_phase_space} 
    Schematic figures in phase space for two different starting points.
    The solid black line is the background trajectory, while the dotted blue line is one of the perturbed trajectories.
    The red and blue arrows show perturbations with $\delta\varphi$ and $\delta\pi$ at the starting point, respectively.
    The vertical dashed and solid lines correspond to $\varphi=\varphi_+$ and $\varphi=\varphi_-$, respectively, which are the boundaries between different stages.
    The four orange dots represent the starting point $(\varphi_\us,\pi_\us)$, boundary points $(\varphi_\pm,\pi_\pm)$, and the end of inflation $(\varphi_\ue,\pi_\ue)$.
    }
\end{figure}

Let us start with the derivation of the $\delta N$ expression.
From the background equation of motion~\eqref{backgroundeom}, we obtain the solution of the scalar field and its velocity in each stage as 
\beae{\label{eq: phipisol}
&
\begin{cases}
\varphi(n)-\varphi_\us=-\frac{1}{3}\left(A_1+\pi_\us\right) e^{-3(n-n_\us)}-A_1 (n-n_\us)+\frac{1}{3}(A_1+\pi_\us)\,,
\\
\pi(n)=(A_1+\pi_\us)e^{-3(n-n_\us)}-A_1\,,
\end{cases}
\quad\hspace{0.17cm}
\text{(1st stage)}\\
&
\begin{cases}
\varphi(n)-\varphi_+=-\frac{1}{3}\left(A_2+\pi_+\right) e^{-3(n-n_+)}-A_2 (n-n_+)+\frac{1}{3}(A_2+\pi_+)\,,
\\
\pi(n)=(A_2+\pi_+)e^{-3(n-n_+)}-A_2\,,
\end{cases}
\text{(2nd stage)}\\
&
\begin{cases}
\varphi(n)-\varphi_-=-\frac{1}{3}\left(A_3+\pi_-\right) e^{-3(n-n_-)}-A_3 (n-n_-)+\frac{1}{3}(A_3+\pi_-)\,,
\\
\pi(n)=(A_3+\pi_-)e^{-3(n-n_-)}-A_3\,,
\end{cases}
\text{(3rd stage)}
}
where the subscript ``$\us$'' denotes a value at the starting point, which provides an initial condition for solutions, and $n_{\us,+,-}$ and $\pi_{\us,+,-}$ are times and velocities when $\varphi=\varphi_{\us,+,-}$, respectively.
Note that $\pi_+$ and $n_+$ are given by the solution for the first stage and $\pi_-$ and $n_-$ are given by the solution for the second stage.
These solutions are uniquely determined by specifying the starting point $(\varphi_\us, \pi_\us)$ in phase space.
The starting point $(\varphi_\us, \pi_\us)$ corresponds to the value of the background fields in phase space at the time when $k=\sigma aH$, where $\sigma<1$ is a constant parameter for the gradient expansion and $k$ is the scale of interest.
In this work, since our main interest is the scales around the dip in the power spectrum, we focus only on scales which exit the Hubble horizon during the first stage.
Hence, we assume that the starting point is set at the first stage.
In Fig.~\ref{schematic_phase_space}, we schematically illustrate the background trajectory (black solid line) and one of the perturbed trajectories (blue dotted line) in phase space.
Time flows from right to left in the figure.
Here, assuming that the background trajectory is on the attractor at the starting point, $(\varphi_\us,\pi_\us)$ is selected accordingly.
Two different choices of the starting point are shown in each panel of the figure.

From the system of equations~\eqref{eq: phipisol}, we obtain the total \efolding number of the background solution by summing contributions from the three stages as
\bae{
N=\sum_{i=1}^{3} N_i\,,
\label{totalefolds}
}
where $N_i$'s denote the time durations for each stage, i.e., $N_1=n_+-n_\us$, $N_2=n_--n_+$, and $N_3=n_\ue-n_-$, which are given by
\beae{
&
    N_1=-\frac{\varphi_{+}-\varphi_{\us}-\frac{1}{3}(A_1+\pi_\us)}{A_1}+\frac{1}{3}W\left(-\frac{A_1+\pi_\us}{A_1}\exp[\frac{3}{A_1}\left(\varphi_{+}-\varphi_{\us}-\frac{1}{3}(A_1+\pi_\us)\right)]\right)\,,\\
&   
    N_2=-\frac{\varphi_{-}-\varphi_{+}-\frac{1}{3}(A_2+\pi_+)}{A_2}+\frac{1}{3}W\left(-\frac{A_2+\pi_+}{A_2}\exp[\frac{3}{A_2}\left(\varphi_{-}-\varphi_{+}-\frac{1}{3}(A_2+\pi_+)\right)]\right)\,,\\
&
    N_3=-\frac{\varphi_{\ue}-\varphi_{-}-\frac{1}{3}(A_3+\pi_-)}{A_3}+\frac{1}{3}W\left(-\frac{A_3+\pi_-}{A_2}\exp[\frac{3}{A_3}\left(\varphi_{\ue}-\varphi_{-}-\frac{1}{3}(A_3+\pi_-)\right)]\right)\,.
}
Here, $W(z)$ denotes the Lambert $W$ function satisfying $z=W(z)e^{W(z)}$, and the subscript ``$\ue$'' denotes a value at the end of inflation.

By perturbing the starting point as $(\varphi_\us,\pi_\us)\rightarrow(\varphi_\us+\delta\varphi, \pi_\us+\delta\pi)$, the solution~\eqref{totalefolds} can describe various perturbed trajectories in the phase space (see Fig.~\ref{schematic_phase_space}).
From the perspective of the $\delta N$ formalism~\cite{Salopek:1990jq,Sasaki:1995aw,Starobinsky:1985ibc,Sasaki:1998ug,Lyth:2004gb,Lee:2005bb,Lyth:2005fi,Abolhasani:2019cqw}, each Hubble patch follows one of these trajectories, and the deviation in the \efolding number from the initial time to the final time, $\delta N$, is equivalent to the curvature perturbation ${\calR}$ in the patch at the final time.
Thus, the curvature perturbation at the end of inflation is expressed as a function of $\delta\varphi$ and $\delta\pi$.

The powerful point of the $\delta N$ formalism is that the curvature perturbation can be treated non-perturbatively as a function of $\delta\varphi$ and $\delta\pi$.
However, as in the usual case, $\delta N$ cannot be written in a simple analytic function.
In order to describe the results analytically and for simplicity, we keep the terms only up to the second order of $\delta\varphi$ and $\delta\pi$, i.e.,
\bae{
{\calR}=\delta N=N_\varphi \delta\varphi+N_\pi \delta\pi+\frac{N_{\varphi\varphi}}{2}\delta\varphi^2+N_{\varphi\pi}\delta\varphi\delta\pi+\frac{N_{\pi\pi}}{2}\delta\pi^2+\calO\pqty{(\delta\varphi,\delta\pi)^3}\,,
\label{deltaN1}
}
where coefficients $N_\varphi$, $N_\pi$, $N_{\varphi\varphi}$, $N_{\varphi\pi}$, and $N_{\pi\pi}$ can be obtained from Eqs.~\eqref{eq: phipisol} and \eqref{totalefolds} (see Appendix~\ref{CoefficientsofdeltaN} for explicit expressions).
Higher-order coefficients can also be obtained by merely expanding Eq.~\eqref{totalefolds} more.
Here we have written down the coarse-grained curvature perturbation in real space, using the $\delta N$ formalism.
This expression can be transformed into the curvature perturbation in the Fourier space for $k\le\sigma a(n_\us)H(n_\us)$.
Since it has the non-linear parts, we should take into account the convolution.
Instead, for simplicity, we impose the so-called single-noise approximation, in which only one mode $k$ is relevant, that is, $\delta\varphi(n_\us(k), \bm{x})=\int \frac{{\rm d}^3p}{(2\pi)^3}\delta(\ln p-\ln k)\delta\varphi_{\bm{p}}(n_\us(k))e^{i\bm{p}\cdot\bm{x}}$, so that the above expression is valid even in the Fourier space without considering the convolution.
This approximation is not generally justified, but can be understood from an intuitive point of view. Namely, as the $\delta N$ is formulated in real space and is computed for a given comoving scale, say $R$, we may assume its crude correspondence to the Fourier mode as $k\sim 1/R$.
It is also widely used in the context of PBH formation from the curvature perturbation (see, e.g., Refs.~\cite{Atal:2019erb,Biagetti:2021eep,Kitajima:2021fpq,Inui:2024fgk,Shimada:2024eec}). 
We follow this treatment for simplicity.

\begin{figure} 
    \centering
    \includegraphics[width=0.8\hsize]{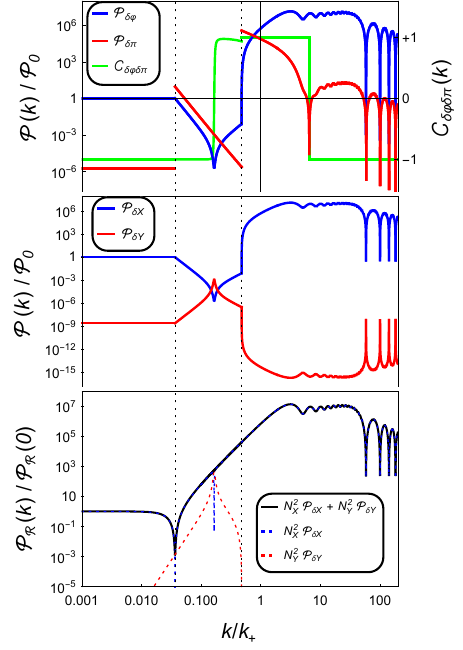}
    \caption{\label{Fig_powerspectrum_XY} 
    \emph{Top panel}: the power spectra of $\delta\varphi$ and $\delta\pi$, and the correlation function of $\delta\varphi$ and $\delta\pi$, defined in Eq.~\eqref{eq: power and correlation}, for the USR model with model parameters shown in the first row of Table~\ref{table1}.
    They are evaluated when $k=\sigma aH$ with $\sigma=0.0367$. The two vertical dotted lines represent $k=\sigma k_+$ and $k=\sigma k_-$.
    The normalization factor ${\cal{P}}_0$ is $H^2/4\Pi^2 \Mpl^2$.
    \emph{Middle panel}: the power spectra of $\delta X$ and $\delta Y$.
    \emph{Bottom panel}: the power spectrum of curvature perturbation $\calR$ calculated by the $\delta N$ formalism at the linear order (black line).
    It is normalised by the power spectrum at $k\to0$.
    The components of Eq.~\eqref{PR}, $N_X^2\calP_{\delta X}$ and $N_Y^2\calP_{\delta Y}$, are also shown (blue and red dotted lines, respectively).
    Thanks to the choice $\sigma=\sigma_\dip=0.0367$ defined in Eq.~\eqref{sigma_dip}, the dip scale is located at $k=\sigma_\dip k_+$.
    Corresponding figures for Starobinsky's linear potential, bump, and step models can be found in Appendix~\ref{figures_other_models}.
    }
\end{figure}

While the $\delta\varphi$ contributions in Eq.~\eqref{deltaN1} dominate in a standard slow-roll model, $\delta\pi$ becomes non-negligible if a non-slow-roll phase is inserted.
To deal with such cases, we introduce new variables $\delta X$ and $\delta Y$ as linear combinations of $\delta\varphi$ and $\delta\pi$ that diagonalise the correlations: 
\be
\delta X=\delta\varphi\, 
\qc
\delta Y=\delta\pi+g \delta\varphi\,,
\ee
where $g$ is a function of the scale of interest $k$ defined by
\be
g\equiv\eval{-\frac{\calC_{\delta\varphi\delta\pi}(k)\sqrt{\calP_{\delta\pi}(k)}}{\sqrt{\calP_{\delta\varphi}(k)}}}_{k=\sigma aH}=\eval{-\Re\qty(\frac{\delta\pi_k}{\delta\varphi_k})}_{k=\sigma aH}
\left(=\frac{\sigma^2} {1+\sigma^2}
\qfor
k<\sigma k_+\right)
\, .
\label{g}
\ee
Here, $\calP_{\delta\varphi}$, $\calP_{\delta\pi}$
and $\calC_{\delta\varphi\delta\pi}$
are the power spectra of $\delta\varphi$ and $\delta\pi$, and the correlation function between $\delta\varphi$ and $\delta\pi$ defined by the mode functions
$\delta\varphi_k$ and $\delta\pi_k$ as\footnote{Here, we suppose that the inflaton perturbation is well classicalised and $\Im(\delta\varphi_k\delta\pi_k^*)$ is negligible. Otherwise, the commutator of $\delta\varphi$ and $\delta\pi$ cannot be neglected and $\calR$'s expansion in these operators in Eq.~\eqref{deltaN1} is not well-defined.}
\be\label{eq: power and correlation}
\calP_{\delta\varphi}(k)\equiv\frac{k^3}{2\Pi^2}|\delta\varphi_k|^2\, 
\qc
\calP_{\delta\pi}(k)\equiv\frac{k^3}{2\Pi^2}|\delta\pi_k|^2\,
\qc
\calC_{\delta\varphi\delta\pi}(k)\equiv\frac{\Re({\delta\varphi_k \delta\pi_k^*})}{|\delta\varphi_k||\delta\pi_k|}\,,
\ee
calculated by the linear theory in the spatially flat gauge.
Here and in the following, the spectra denoted by the symbol $\calP$ are those obtained in linear theory as given above, unless otherwise stated.
In the top panel of Fig.~\ref{Fig_powerspectrum_XY}, we show these power spectra for the USR model evaluated at the time when $k=\sigma aH$ for each mode, where we chose $\sigma=0.0367$.
This value corresponds to $\sigma_\dip$, which will be introduced later.

In the last equality in Eq.~\eqref{g}, since we are interested in the dip scale, we assumed a case where the starting point is located in the first stage ($k<\sigma k_+$).
One can check that the above definition of $g$ indeed cancels the correlation between $\delta X$ and $\delta Y$ (i.e., $\calC_{\delta X\delta Y}=0$) and minimises the power spectrum of $\delta Y$.
Their auto power spectra are given by
\beae{
&
\calP_{\delta X}(k)=\calP_{\delta\varphi}(k)\left(=(1+\sigma^2)\left(\frac{H}{2\Pi \Mpl}\right)^2 
\qfor k<\sigma k_+ \right)
\,,
\\
&
\calP_{\delta Y}(k)=\frac{1}{\calP_{\delta\varphi}(k)}\left(\frac{k}{aH}\right)^6\left(\frac{H}{2\Pi \Mpl}\right)^4\left(=\frac{\sigma^6}{1+\sigma^2}\left(\frac{H}{2\Pi \Mpl}\right)^2 
\qfor k<\sigma k_+ \right)
\label{powerspectrum_XY}
\,.
}
These spectra are exhibited in the middle panel of Fig.~\ref{Fig_powerspectrum_XY} for the USR model.
Note that $\calP_{\delta\varphi}(k)\calP_{\delta Y}(k)\propto\sigma^6$ is independent of $k$ and hence the power spectrum of $\delta Y$ is always suppressed by $\sigma^6$, except for the value of $k$ at which $\calP_{\delta\varphi}(k)$ is suppressed (i.e., at a dip of $\calP_{\delta\varphi}$).
This makes the variable $\delta Y$ particularly convenient, in comparison with $\delta\pi$ whose power is not necessarily suppressed compared to that of $\delta \varphi$ (see the top panel of Fig.~\ref{Fig_powerspectrum_XY}).

\begin{figure} 
    \centering
    \includegraphics[width=0.9\hsize]{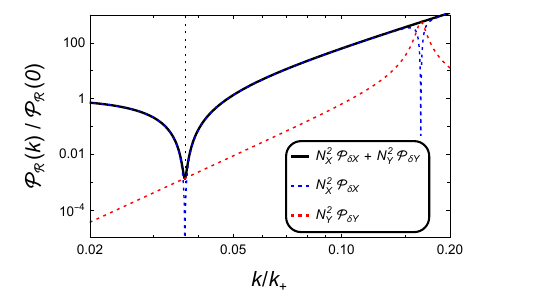}
    \caption{\label{Fig_powerspectrum_Zoim_in} 
    Enlarged version of the bottom panel of Fig.~\ref{Fig_powerspectrum_XY} around the dip scale.
    }
\end{figure}

With these new variables, we can rewrite Eq.~\eqref{deltaN1} as 
\be
{\calR}=N_X \delta X+N_Y \delta Y+\frac{N_{XX}}{2}\delta X^2+N_{XY}\delta X\delta Y+\frac{N_{YY}}{2}\delta Y^2+\calO\pqty{(\delta X,\delta Y)^3}
\label{deltaN2}\,,
\ee
where the coefficients are related to those in Eq.~\eqref{deltaN1} as
\bege{
\label{eq:NN relations}
N_X=N_{\varphi}-g N_\pi\, \qc 
N_Y=N_\pi
\\
N_{XX}=N_{\varphi\varphi}-2g N_{\varphi\pi}+g^2 N_{\pi\pi}\, \qc 
N_{XY}=N_{\varphi\pi}-g N_{\pi\pi}\, \qc 
N_{YY}=N_{\pi\pi}\,.
}
These coefficients can also be derived directly by the perturbative expansion of Eq.~\eqref{totalefolds} in terms of $\delta X$ and $\delta Y$.

\section{Linear analysis and the dip condition}
\label{linearanalysis}
In this section, we only focus on linear theory. We study how the result from the $\delta N$ formalism corresponds to that obtained in linear perturbation theory.
The discussion here aims to find a condition for the appearance of a dip, ``the dip condition''.

At the linear level, we can drop the quadratic terms in Eq.~\eqref{deltaN2} and obtain the power spectrum of $\calR$ in a simple form,
\be
\calP_{\calR}=N_{X}^2\calP_{\delta X}+N_{Y}^2\calP_{\delta Y}\,,
\label{PR}
\ee
where $\calP_{\delta X}$ and $\calP_{\delta Y}$ are the power spectra of $\delta X$ and $\delta Y$, respectively.
Here we used the fact that the correlation function between $\delta X$ and $\delta Y$ is zero.

\begin{figure}
    \centering
    \begin{tabular}{c}
        \begin{minipage}{0.95\hsize}
            \centering
            \includegraphics[width=
            0.9\hsize]{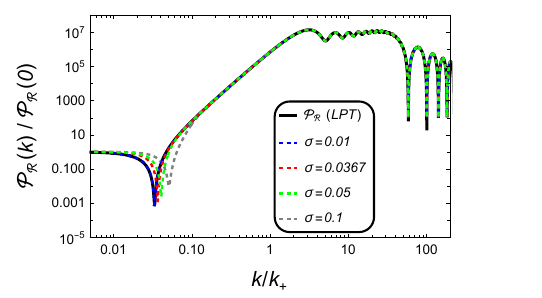}
        \end{minipage}
        \\
        \\
        \begin{minipage}{0.95\hsize}
            \centering
            \includegraphics[width=
            0.9\hsize]{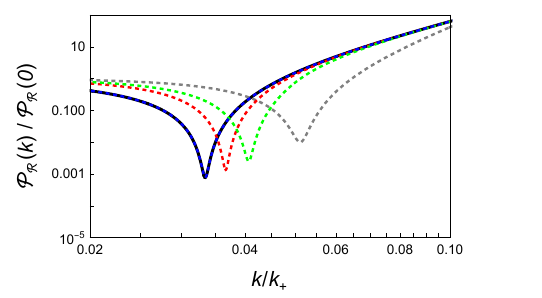}
        \end{minipage}
    \end{tabular}
    \caption{\label{powerspectrumsigma}  
    \emph{Upper}: the power spectrum of curvature perturbation $\calR$ for the USR model (see Table~\ref{table1} in Sec.~\ref{Appendixpowerspectrum} for detailed parameters) calculated by linear perturbation theory (black solid line) and $\delta N$ formalism with various gradient expansion parameter $\sigma$ (coloured dotted lines).
    These are normalised by the power spectrum at $k=0$.
    \emph{Lower}: enlarged version of the power spectrum around the dip scale.
    Corresponding figures for Starobinsky's linear potential, bump, and step models can be found in Appendix~\ref{figures_other_models}.}
\end{figure}

Let us compare the result obtained from Eq.~\eqref{PR} with the result obtained from linear perturbation theory.
In Fig.~\ref{powerspectrumsigma}, we show a comparison between the power spectrum calculated from the linear perturbation theory (black solid line) and the results calculated using Eq.~\eqref{PR} for the USR model with various values of the gradient expansion parameter $\sigma$ (dotted lines).
The detailed calculation for the linear perturbation theory and the used model parameters can be found in Sec.~\ref{Appendixpowerspectrum}.
The results from the two calculations agree to a considerable degree of accuracy for a sufficiently small $\sigma$ value.\footnote{In this study, we ignore the spatial gradient corrections which become important in the slow-roll violating model~\cite{Leach:2001zf}.
Readers interested in an analysis including such corrections should refer to~\cite{Takamizu:2010xy,Naruko:2012fe} for pioneer works and \cite{Jackson:2023obv,Artigas:2024xhc} for recent development.}
Even with larger $\sigma$ and hence in a worse approximation of gradient expansion, most parts of the power spectrum generally agree with the result from linear perturbation theory.
However, as $\sigma$ increases, the calculation result around the dip becomes misaligned (see the lower panel of Fig.~\ref{powerspectrumsigma}).
Therefore, $\sigma$ must be sufficiently small to accurately reproduce the power spectrum around the dip scale.

%

Let us have a look at each of the contributions from the two terms on the right-hand side of Eq.~\eqref{PR}.
In the bottom panel of Fig.~\ref{Fig_powerspectrum_XY} (see also Fig.~\ref{Fig_powerspectrum_Zoim_in}), we show the contributions from each term in Eq.~\eqref{PR} (blue and red dotted lines).
The power spectrum of the curvature perturbation is dominated by $N_{X}^2\calP_{\delta X}$ for most of the scales except for two scales.\footnote{If we take the limit of $\sigma\to0$, two scales would coincide with each other.
However, this is the same as solving for all-time regimes with a linear perturbation theory, which is not appropriate for our purpose.}
One is a scale around $k/k_+=0.166$ at which $\eval{\calP_{\delta\varphi}}_{k=\sigma aH}$ has a dip so that $\calP_{\delta X}$ has a minimum and $\calP_{\delta Y}$ has a maximum.
This scale, however, is not our current interest in this paper, as the total power spectrum has no feature there.
The other is the dip scale $k_\dip\simeq0.0367k_+$ and our current interest.
From the figure, at the dip scale, the contribution from $\delta X$ vanishes and $\delta Y$ 
dominates the contribution to $\calR$. 
Note that this is not because $\calP_{\delta X}$ becomes smaller but because the coefficient $N_X$ vanishes at the dip scale, (see the first relation in Eq.~\eqref{eq:NN relations}) i.e.,
\be
N_X=N_\varphi-gN_\pi=0 ~\Leftrightarrow~
\frac{N_\pi}{N_\varphi}=1/g
\quad\text{at}~ k=k_\dip 
\,.
\label{kdipcondition}
\ee
We call the above equation \emph{the dip condition} in the following.\footnote{Recently, it has been shown that the dip does not appear in the overshoot case, where the inflaton turns back due to the potential and its velocity changes the sign~\cite{Wang:2024wxq,Briaud:2025hra}. 
Our analysis does not include such cases.
We leave it for future work to see how the dip condition changes in these situations.}
In fact, the middle panel of Fig.~\ref{Fig_powerspectrum_XY} shows that $\calP_{\delta X}$ does not decrease at the dip scale at all.
Here, let us consider why a particular scale satisfying the dip condition appears in the transient non-slow-roll models.
In the slow-roll stage, the $\delta \pi$ displacement quickly converges to the attractor trajectory, and hence the effect from $\delta\pi$ on $\delta N$ should be negligibly small.
In fact, the typical perturbation amplitude of $\delta\pi$ is suppressed compared to that of $\delta\varphi$ as $\calP_{\delta\pi}/\calP_{\delta\varphi}=\sigma^4/(1+\sigma^2)$ during the first slow-roll phase, and their coefficients are comparable with each other for $k\ll k_+$:
\be
\frac{N_\varphi}{N_\pi}=3 \quad \text{for}\quad\varphi_\us\to+\infty\,,
\ee
where we used Eqs.~\eqref{N_varphi_first} and \eqref{N_pi_first} in Appendix with the limit $N_1\rightarrow+\infty$.
On the other hand, the $\delta \pi$ displacement becomes crucial when the starting point $\varphi_\us$ is close to $\varphi_+$.
This is because the non-slow-roll stage at $\varphi<\varphi_+$ begins before the perturbed trajectory with an initial non-vanishing $\delta\pi$ converges to the slow-roll attractor trajectory.
One can observe this fact from Fig.~\ref{schematic_phase_space} by comparing the two panels.
Indeed, it is shown as follows that $N_\varphi/N_\pi$ at $\varphi_\us=\varphi_+$ is much smaller than unity to overcome the suppression of the $\delta\pi$'s power spectrum as long as the significant amplification is realised at the peak scale in the power spectrum of the curvature perturbation.
That is, one finds
\be
\frac{N_\varphi}{N_\pi}=\frac{3A_3}{A_1}\frac{A_1+A_2(e^{3N_2}-1)}{A_1+A_3(e^{3N_2}-1)}=\frac{3A_3\pi_-}{\bar{A}\pi_+}=3\left(\frac{\calP_\calR^{(\SR)}}{\calP_\calR^{(\peak)}}\right)^{1/2}\ll1
\quad\text{at}\quad \varphi_\us=\varphi_+\,,
\ee
where the first equality follows from the explicit results of the $\delta N$ calculation (see Eqs.~\eqref{N_varphi_first} and \eqref{N_pi_first} in Appendix) with $N_1=0$, which corresponds to $k=\sigma k_+$ ($\varphi_\us=\varphi_+$),
 $\bar{A}$ is the weighted average of $A_1$ and $A_3$ defined in Eq.~\eqref{weighted_average},
and $\pi_+$ and $\pi_-$ are the velocity evaluated at $\varphi=\varphi_+$ 
and $\varphi_-$, respectively, i.e.,
\be
\pi_+=-A_1\, 
\qc
\pi_-=(A_2-A_1)e^{-3N_2}-A_2\,.
\ee
The third equality follows from the approximated power spectrum at the peak given by~\eqref{approximated_power}.
Therefore, according to the intermediate value theorem, $N_\varphi/N_\pi$ at a certain $\varphi_\us$ ($\varphi_+\le\varphi_\us<+\infty$)  coincides with some small constant $g$ defined in Eq.~\eqref{g} as long as one chooses $\sigma$ as $\sigma^2\simeq g\ge \eval{(N_\varphi/N_\pi)}_{\varphi_\us=\varphi_+}$ so that the dip condition~\eqref{kdipcondition} is satisfied in the first slow-roll stage.
From the above discussion, for models that include one/multiple periods during which the inflaton loses its velocity, the dip scale at which $N_X=0$ is generally satisfied appears.
This result is one of the main messages of this paper.

\begin{figure}
    \centering
    \begin{tabular}{c}
        \begin{minipage}{0.95\hsize}
            \centering
            \includegraphics[width=
            0.95\hsize]{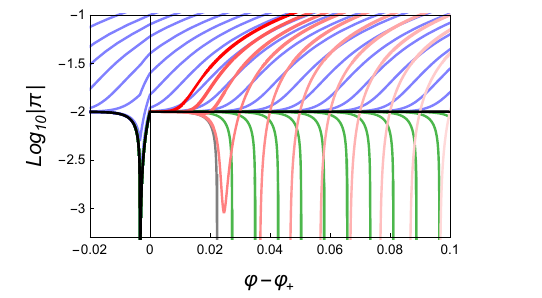}
        \end{minipage}
        \\
        \\
        \begin{minipage}{0.95\hsize}
            \centering
            \includegraphics[width=
            0.95\hsize]{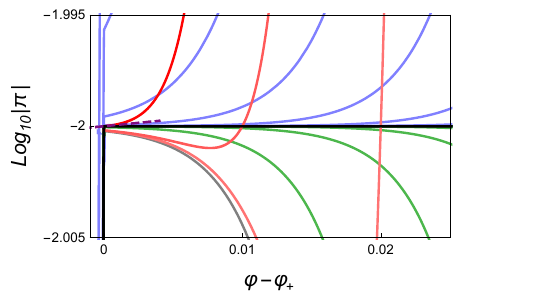}
        \end{minipage}
    \end{tabular}
    \caption{\label{phasespacediagram}  
    \emph{Upper panel}: phase-space diagram in the USR case.
    The black line demonstrates the background trajectory while  
    the blue and green lines denote 
    perturbed trajectories with various initial conditions. 
    The gray line is the critical trajectory above which the inflaton can overcome the USR stage in a finite time.
    The red lines show the equal $N$ contours with interval $\Delta N=1$.
    \emph{Lower panel}: a zoom-in version around the end of the first stage. The purple dotted line denotes the displacement in the $X$-direction for $\sigma=\sigma_{\rm dip}=0.0367$.}
\end{figure}

For a visual understanding of the above description, a phase space diagram in the case of the USR model is depicted in Fig.~\ref{phasespacediagram}.
The black line denotes the trajectory of the background, and since $\pi$ is always negative in the present setup, the time evolution is from right to left in the diagram.
The background trajectory is on the slow-roll attractor from the beginning in the first slow-roll stage, the inflaton then slows down in the \ac{USR} stage, and the trajectory finally asymptotes to the second slow-roll attractor.
In Fig.~\ref{phasespacediagram}, we also illustrate trajectories with different initial conditions with blue, green, and gray lines.
These trajectories are not on the attractor at the beginning, but asymptotically approach the slow-roll attractor with time evolution.
The red lines connect points in phase space that have equal \efolding numbers to reach a certain value $\varphi_\ue$ in the second slow-roll stage, i.e., the backward-$N$ contours.
From the figure, it can be seen that in the region sufficiently far from $\varphi_+$ in the positive direction, the backward-$N$ contours intersect the background trajectory with a significant angle, whereas in the vicinity of $\varphi_+$, the $N$ constant lines intersect the background trajectory in an almost horizontal direction (see the lower panel of Fig.~\ref{phasespacediagram}).
This fact indicates that whether $N_\varphi$ or $N_\pi$ has a major effect on $\delta N$ is flipped between the two regions, i.e. $N_\varphi/N_\pi$ is large enough in the region $\varphi\gg\varphi_+$, while it is small in the region $\varphi\sim\varphi_+$.
This is consistent with the explanation given in the previous paragraph.\footnote{While our explicit expressions in the text rely on the slow-roll limit (in particular $\abs{\eta}\coloneqq\abs{(\dv*{\pi}{n})/(\dv*{\varphi}{n})}\ll1$), our discussion would be generalised to a non-negligible and non-constant $\eta$ case. Here, the tilt of the background trajectory is given by $\eta$. Similarly to the slow-roll limit, the tilt of the equal $N$ contours, $-N_\varphi/N_\pi$, is expected to form a certain $\calO(1)$ angle with the background trajectory, i.e., $N_\varphi/N_\pi=-\eta+c$ with an $\calO(1)$ constant $c$ for $k\ll k_+$, while $N_\varphi/N_\pi$ asymptotes to $-\eta$ at $\varphi_\us=\varphi_+$.
On the other hand, by expanding the general solution for the Mukhanov--Sasaki variable, one finds that $g\simeq-\eta+\sigma^2$ at the leading order in $\sigma$ and $\eta$. Therefore, again according to the intermediate value theorem, there is a point which satisfies the dip condition for an appropriately chosen $\sigma\ll1$.}

The dip condition~\eqref{kdipcondition}, i.e., $N_X=0$ implies that near the dip scale, the effect of $\delta Y$ cannot be ignored at all, even though the power spectrum of $\delta Y$ is suppressed by the gradient expansion parameter $\sigma$.
Let us evaluate the power spectrum at the dip scale by using the dip condition.
For simplicity, we choose the gradient expansion parameter $\sigma$ as $k_\dip=\sigma k_++\varepsilon$ where $\varepsilon$ is a negligibly small positive constant.
In this case, $\sigma$ is given by a solution of
\be
\sigma_{\rm dip}^2\simeq g=\eval{\frac{N_\varphi}{N_\pi}}_{\varphi_\us=\varphi_+}=\frac{3A_3\pi_-}{\bar{A}\pi_+}\,.
\label{sigma_dip}
\ee
The dip scale, $k_\dip$, is given by substituting $\sigma$ into $k_\dip=\sigma k_+$.
The choice of $\sigma$ in this manner simplifies the later analysis, as the starting point $\varphi_\us$ corresponding to the dip scale is located during the first stage, $\varphi_\us>\varphi_+$.\footnote{
As we see in Fig.~\ref{powerspectrumsigma}, the smaller the $\sigma$ is, the more accurately the linear perturbation theory result is reproduced.
One may wonder why we do not choose an even smaller $\sigma$.
If $\sigma$ is set too small, as in  $k_\dip>\sigma k_+$, then the starting point is no longer in the first stage.
In this case, the power spectra of $\delta X$ and $\delta Y$ become cumbersome and the intrinsic non-Gaussianities, which are the non-Gaussianities of $\delta X$ and $\delta Y$, cannot be ignored~\cite{Ballesteros:2024pbe}.
This would negate the advantages of the $\delta N$ formalism, which should be easy to compute.
In this paper, therefore, the $\sigma$ is taken to be large enough to satisfy $k_\dip<\sigma k_+$, reluctantly accepting that the deviation due to higher-order corrections for gradient expansion is missed.}
With this choice, the power spectrum of the curvature perturbation at the dip scale in linear perturbation can be easily computed as
\be
{\cal{P}}_{{\cal{R}}}^{({\rm dip})}=N_{Y}^2{\cal{P}}_{\delta Y}\simeq \frac{3A_3}{A_1^3}\frac{A_1+A_2(e^{3N_2}-1)}{A_1+A_3(e^{3N_2}-1)}\frac{H^2}{4\Pi^2 \Mpl^2}\simeq \frac{3A_3\pi_-}{\bar{A}\pi_+}
\calP_\calR^{(\SR)}\,,
\label{power_at_dip}
\ee
where, in the second approximate equality, we used $N_Y=N_\pi$ and Eq.~\eqref{N_pi_first} with $N_1=0$ for rewriting $N_Y$, and Eq.~\eqref{powerspectrum_XY} and Eq.~\eqref{sigma_dip} for rewriting $\calP_{\delta Y}$.
As we have seen in Sec.~\ref{Appendixpowerspectrum}, the power spectrum at the peak scale is approximately given by Eq.~\eqref{approximated_power}.
Thus, we find that the depth of the dip is strongly correlated to the amplification of the power spectrum at the peak. Roughly speaking, we have $\text{(deepness of dip)}^{-2}\simeq \text{(amplification of peak)}$ (see Ref.~\cite{Briaud:2025hra} for a detailed discussion in linear perturbation theory).
At the linear level, the power spectrum of the curvature perturbation has a nonzero value at the dip scale due to the existence of the $\delta Y$ contribution as seen above.
Note, however, that if the non-linear effect is included, $\delta X$ also contributes to $\calR$ at higher orders.
Therefore, $\calR$ must be treated carefully as a bivariate function around the dip scale.

\section{Non-linearity around the dip}
\label{nonlinearity}
In the previous section, we showed that the coefficient $N_X$ vanishes and 
the power spectrum of the curvature perturbation is supported by $\delta Y$ at the dip $k=k_\dip$.
In other words, the dominant contribution to the curvature perturbation is not from $\delta X$ but from $\delta Y$, at least in linear perturbation.
In this section, let us consider the effects of non-linearity.
More specifically, we want to know whether the dominant contribution to $\calR$ is still from $\delta Y$ or $\delta X^n$ ($n\ge2$).

We first briefly discuss and illustrate the \ac{PDF},  $P[{\calR}]$,  of the curvature perturbation based on numerical calculations.
Let us review how the PDF of the curvature perturbation is calculated.
We already know the curvature perturbation $\calR$ as a function of $\delta X$ and $\delta Y$ as a result of the $\delta N$ formalism.
In this case, the PDF of $\calR$ is converted from the PDF of $\delta X$ and $\delta Y$, $P[\delta X,\delta Y]$, by the probability conservation law,
\be
P[\calR]=\dv{}{\calR}\int_{\Omega}P[\delta X,\delta Y]\dd{\delta X} \dd{\delta Y}
=\int_{-\infty}^{+\infty}\int_{-\infty}^{+\infty}
P[\delta X,\delta Y]\delta \qty({\calR}-\delta N(\delta X,\delta Y))
\dd{\delta X} \dd{\delta Y}
\label{PDFconversion}
\,,
\ee
where the integral region $\Omega$ is defined by
\be   \Omega\coloneqq\Bqty{(\delta X,\delta Y)\mid-\infty<\delta N(\delta X, \delta Y)\leq{\calR}}\,.
\ee
By construction, $\delta X$ and $\delta Y$ are uncorrelated with each other, so that $P[\delta X,\delta Y]$ is provided by the product of these independent \acp{PDF}.
As we assume that the initial condition is given in the first stage, the non-Gaussianities of $\delta X$ and $\delta Y$ can be ignored as they are of high order in slow-roll corrections. Then their PDFs are Gaussian with high accuracy with variances $\calP_{\delta X}$ and $\calP_{\delta Y}$, respectively.
Thus, 
\be
P[\delta X, \delta Y]=\frac{1}{\sqrt{2\Pi {\cal{P}}_{\delta X}}}\frac{1}{\sqrt{2\Pi \calP_{\delta Y}}}\exp\left(-\frac{\delta X^2}{2\calP_{\delta X}}\right)\exp\left(-\frac{\delta Y^2}{2\calP_{\delta Y}}\right)\,.
\ee

\begin{figure}
    \centering
    \begin{tabular}{c}
        \begin{minipage}{0.95\hsize}
            \centering
            \includegraphics[width=0.95\hsize]{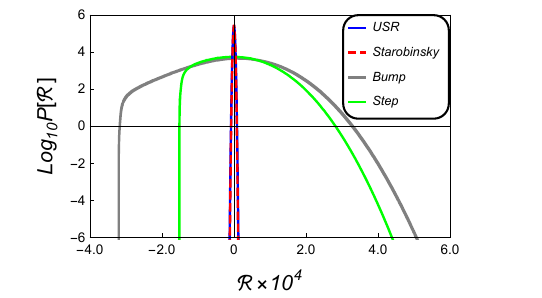}
        \end{minipage}
        \\
        \\
        \begin{minipage}{0.95\hsize}
            \centering
            \includegraphics[width=0.95\hsize]{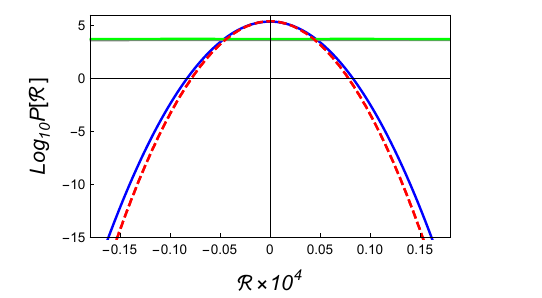}
        \end{minipage}
    \end{tabular}
    \caption{\label{dipPDF} 
    \emph{Upper panel}: the \ac{PDF} of the curvature perturbation for the four different models; USR model (blue), Starobinsky's linear potential model (red), bump model (gray), and step model (green).
    The detailed parameters are shown in Table~\ref{table1}.
    \emph{Lower panel}: a zoom-in version around ${\calR}=0$.}
\end{figure}

From the explicit form of the \efolding number $N$ as a function of $\delta\varphi$ and $\delta\pi$, Eq.~\eqref{totalefolds}, we can easily derive $\delta N=N-\ev{N}$ as a function of $\delta X$ and $\delta Y$ where $\langle~ \rangle$ denotes the expectation value.
Using it and the PDF conversion formula~\eqref{PDFconversion}, we illustrate the \acp{PDF} of the curvature perturbation in Fig.~\ref{dipPDF} for the four models given in Table~\ref{table1}.
Here we focus on the dip scale defined by Eq.~\eqref{kdipcondition}. 
As we expected, we can see that the non-linear effects are pronounced in the bump and step models.

Using the resultant \acp{PDF}, we also show the variance $\ev{{\calR}^2}$, the skewness $\ev{{\calR}^3}/\ev{{\calR}^2}^{3/2}$, and the kurtosis $(\ev{{\calR}^4}/\ev{{\calR}^2}^{2}-3)$ in Table~\ref{extable}.
The Hubble parameter $H$ for each model is chosen in such a way that the power spectrum of all four models matches $2\times10^{-9}$ in the long wavelength limit.
Hence, at the linear level, the four models show almost the same order of magnitude of the power spectrum at the dip scale, $\calP_{\calR}(k_\dip)\sim10^{-12}$ (see Fig.~\ref{powerspectrum}).
However, the variances based on the non-linearly computed PDFs are very different among the models.
This is because, in the bump and step models, the non-linear contributions of $\delta X$ and $\delta Y$ to $\mathcal{R}$ become dominant over the linear ones and lead to much larger variances of the curvature perturbation.
As further evidence, the skewness and kurtosis are also larger. Thus, the deviation from the Gaussian distribution is prominent in those two models.

\begin{table}
    \renewcommand{\arraystretch}{1.3}
    \centering
    \begin{tabular}{c @{\quad} c @{\quad} c @{\quad} c} 
        \toprule
        Model & Variance & Skewness & Kurtosis \\ \hline 
        USR & $2.79\times10^{-12}$ & $-0.00115$ & $-0.00172$\\
        Starobinsky linear & $2.48\times10^{-12}$  & $-0.000147$ & $-5.05\times10^{-6}$ \\
        Bump & $8.08\times10^{-9}$ & $-0.235$ & $0.0318$ \\
        Step & $4.57\times10^{-9}$ & $0.221$ & $-0.363$ \\
        \bottomrule
    \end{tabular}
    \caption{Statistical quantities for four models}
    \label{extable}
\end{table}

For the bump and step models, a sharp cutoff can be seen on the left edge of the PDF in Fig.~\ref{dipPDF}.
The same feature was observed in our previous study~\cite{Kawaguchi:2023mgk}, in which highly asymmetric PDFs were discussed, where the probability of finding fluctuations that give $\delta N<0$ is negligible.
This feature reflects the fact that deviations along the $X$-direction from the background trajectory, as shown by the purple dashed line in the lower panel of Fig.~\ref{phasespacediagram}, always give rise to $\delta N>0$ at the dip.
Intuitively, this may be explained as follows.
If $\delta X (=\delta\varphi)>0$, although the inflaton obtains a faster initial velocity, the distance to $\varphi_e$ becomes longer, which leads to a larger number of \efolds.
On the other hand, if $\delta X<0$, the distance to $\varphi_e$ becomes shorter, but the inflaton has a slower initial velocity, which also leads to a larger number of \efolds.
As a result, the trajectory with $\delta X=0$ reaches $\varphi_e$ earliest at the dip.
In other words, the \efolding number to the end of inflation is not shortened if $\delta X$ varies.
Unlike the previous study~\cite{Kawaguchi:2023mgk}, this study incorporates the effect of $\delta\pi$.
Thanks to the presence of the $\delta Y$ contribution, the PDF does not go to zero even in the large negative region of $\calR$.
However, since the main contribution from $\delta X$ is absent, the value of the PDF decreases sharply and appears to show a cutoff.

So far, we have shown from the numerical calculation that significant deviations from the Gaussian distribution are found for the bump and step models.
As a next step, we consider how to analytically evaluate the non-linearity of the curvature perturbation.
Unfortunately, it is difficult to obtain an analytic expression of the non-linear PDF in Eq.~\eqref{PDFconversion}.
Hence, we resort to the series expansion of the curvature perturbation~\eqref{deltaN2}.
Furthermore, since $\calP_{\delta X}\gg \calP_{\delta Y}$, the dominant contribution in the quadratic terms in Eq.~\eqref{deltaN2} comes from $\delta X^2$. Hence, we ignore the other terms, i.e., we have
\be
\calR\simeq N_Y \delta Y+\frac{N_{XX}}{2}\left(\delta X^2-\langle\delta X^2 \rangle\right)+\cdots\,.
\label{reducedR}
\ee
Note that we dropped the term linear in $\delta X$ due to the dip condition, Eq.~\eqref{kdipcondition}, and we shift $\calR$ by a constant so that its expectation value vanishes, $\langle\calR\rangle=0$.
From Eq.~\eqref{reducedR}, we can calculate the variance of $\calR$ as 
\be
\langle\calR^2 \rangle=\int_{-\infty}^{+\infty}{\calR}^2P[\calR]\dd{\calR}\simeq N_Y^2\calP_{\delta Y}+\frac{1}{2}N_{XX}^2\calP_{\delta X}^2+\cdots\,,
\label{variance_R}
\ee
where we assume that $\delta X$ and $\delta Y$ follow the exact Gaussian distribution with variance $\calP_{\delta X}$ and $\calP_{\delta Y}$, respectively.

Let us introduce a parameter which expresses the degree of non-linearity at the dip,
\be
\Delta\equiv\frac{N_{XX}^2\calP_{\delta X}^2(k_\dip)}{N_Y^2 \calP_{\delta Y}(k_\dip)}\simeq
\frac{N_{XX}^2\calP_{\delta \varphi}(k_\dip)}{N_Y^2 \sigma^6}\simeq
\frac{(2N_{\varphi\pi}-gN_{\pi\pi})^2}{N_\varphi N_\pi}\calP_{\delta \varphi}(k_\dip)\,,
\label{Deltadef}\ee
where $k_\dip<\sigma k_+$ is assumed, and we used Eqs.~\eqref{g}, \eqref{powerspectrum_XY} and \eqref{kdipcondition}.
If $\Delta\ll1$, the curvature perturbation is dominated by the $N_Y \delta Y$ contribution and hence has a Gaussian PDF.
On the other hand, if $\Delta\gg1$, the curvature perturbation becomes highly non-Gaussian.
Below, we will discuss what types of models have large values of $\Delta$.

With the above choice~\eqref{sigma_dip}, the non-linearity parameter $\Delta$ can be expressed in general form as
\be
\Delta \simeq\frac{12A_3^2}{\bar{A}^2}\left(-\frac{\bar{A}}{A_3}+\frac{\pi_-}{\pi_+}-\frac{\pi_+ + \pi_-}{2\pi_- e^{3N_2}}\right)^2 \sqrt{\calP^{(\SR)}_{\calR}\calP^{(\peak)}_{\calR}}\,,
\label{generalDelta}
\ee
where we used explicit results of $\delta N$ calculation with $N_1=0$ to derive the above equation (see Eqs.~\eqref{N_varphi_first}--\eqref{N_pipi_first} in Appendix~\ref{CoefficientsofdeltaN}) and $\calP^{(\SR)}_{\calR}$ and $\calP^{(\peak)}_{\calR}$ are approximate power spectra, which are defined by Eq.~\eqref{approximated_power}.
Since we assume that the Hubble parameter is a constant, $\calP^{(\SR)}_{\calR}$ can be replaced by the value at the CMB scale, i.e., $\calP^{(\SR)}_{\calR}\simeq2\times 10^{-9}$.
In addition, the power spectrum at the peak scale, $\calP^{(\peak)}_{\calR}$, is at most $10^{-2}$ to avoid PBH overproduction.
Therefore the non-linearity parameter $\Delta$ is suppressed by the factor $(\calP^{(\SR)}_{\calR}\calP^{(\peak)}_{\calR})^{1/2}\lesssim 10^{-5.5}$.
It can be seen that if one wants to build a model in which the non-linear effects are dominant, the way to overcome the suppression is to increase the inside of the parentheses in Eq.~\eqref{generalDelta}.
Let us take a look at the typical cases of the three  models:
\begin{itemize}
\item USR model ($\pi_-=\pi_+ e^{-3N_2}$, $N_2\gg1$, $A_3\gg|\pi_-|$)
\be
\bar{A}\simeq A_3\,\qc
g\simeq3 e^{-3N_2}=\frac{3\pi_-}{\pi_+}\, 
\qc\Delta\simeq27\sqrt{\calP^{(\SR)}_{\calR}\calP^{(\peak)}_{\calR}}\ll1 \,.
\ee
\item Starobinsky's linear potential model ($N_2=0$, $\pi_+=\pi_-$, $A_3\ll A_1$)
\be
\bar{A}\simeq A_1\,\qc
g\simeq\frac{3A_3}{A_1}\, 
\qc
\Delta\simeq12\sqrt{\calP^{(\SR)}_{\calR}\calP^{(\peak)}_{\calR}}\ll1\,.
\ee
\item Bump/Step models ($N_2\ll1$, $|A_2|\gg A_1$, $|\pi_-|\ll A_1,A_3$)
\be
\bar{A}\simeq A_1\,\qc
g\simeq\frac{3A_3\pi_-}{A_1\pi_+}\, 
\qc
\Delta \simeq 3\left(\frac{A_3}{A_1}\right)^2\left(\frac{\pi_+}{\pi_-}\right)^2\sqrt{\calP^{(\SR)}_{\calR}\calP^{(\peak)}_{\calR}}=3\frac{A_3}{A_1}\left(\frac{\pi_+}{\pi_-}\right)^3\calP^{(\SR)}_{\calR}\,.
\label{Delta_step}
\ee
The last equality implies $\Delta$ can be larger than unity if $(A_3/A_1)^{1/3}(\pi_+/\pi_-)\gtrsim 10^{3}$.
Therefore, in the bump or step model, the non-linearity can dominate at the dip scale. Hence,  we cannot trust the linear perturbation theory result there.
Indeed, for the model parameters we used, the amplification factor at the peak is around $10^7$, or $(\pi_+/\pi_-)^2\sim10^{7}$, and $A_1=A_3$, which means that the non-linearity comes into play.
In these cases, omitting the contributions from higher-order terms, the variance of the curvature perturbation~\eqref{variance_R} is determined by the second-order term, $(\Delta/2)\calP^{(\dip)}_{\calR}$, and can be evaluated as
\be
\langle\calR^2 \rangle^{(\rm dip)}\simeq\frac{\Delta}{2}\calP_\calR^{(\dip)}\simeq
\frac{9(A_3\pi_+)^2}{2(A_1\pi_-)^2}\left(\calP_\calR^{(\SR)}\right)^2\simeq\frac{9}{2}\left(\frac{A_3}{A_1}\right)^4{\calP^{(\peak)}_{\calR}}\calP^{(\SR)}_{\calR}\,,
\label{variance_around_dip}
\ee
where we used Eqs.~\eqref{approximated_power} and \eqref{power_at_dip}.
This can be much larger than one from the linear level calculation~\eqref{power_at_dip}.\footnote{Note that this result is obtained by using the single noise approximation.
In Ref.~\cite{Franciolini:2023agm}, they calculate the one-loop correction to the dip from all scales, including the peak scale, in the case of the USR model.
They conclude that the dip becomes shallower even in the USR model.
Therefore, we can naively expect that the one-loop correction from the peak scale is not negligible for the bump/step models and may change the result we derived.
We leave this issue for future work.}
Note that $\Delta$ is proportional to $(\pi_+/\pi_-)^3$ as well as to $A_3/A_1$.
Since the amplification of the power spectrum at the peak is given by $(A_1\pi_+)^2/(A_3\pi_-)^2$ as in Eq.~\eqref{approximated_power}, $\pi_+/\pi_-$ must scale as $A_3/A_1$ to sustain the amplification, $(A_1/A_3)(\pi_+/\pi_-)\sim{\text{const}}$. Thus, a larger $A_3/A_1$ leads to an even larger value of $\Delta$.
This implies that the potential slope $A_3$ after the bump/step stage plays a crucial role in the non-linearity at the dip. A sharp change, i.e, $|A_3|\gg|A_1|$, produces the largest non-linearity in the bump/step models.
\end{itemize}
\begin{table}
    \renewcommand{\arraystretch}{1.3}
    \centering
    \begin{tabular}{c @{\quad} c @{\quad} c} 
        \toprule
        Model &$\sigma_\dip$ & $\Delta$  \\ \hline 
        USR & 0.0367 &$1.20\times10^{-4}$ \\
        Starobinsky linear & 0.0347 &$6.00\times10^{-5}$ \\
        Bump & 0.0339 & 38.4 \\
        Step & 0.0277 &357 \\
        \bottomrule
    \end{tabular}
    \caption{$\sigma_{\dip}$ and $\Delta$ for four models}
    \label{table3}
\end{table}
The values of $\sigma_{\dip}$ and $\Delta$ are shown in Table~\ref{table3} for the specific models presented in Table~\ref{table1}.
We confirm that these results are consistent with the approximate equations obtained above.

As we have discussed, only the bump and step cases can realise the non-linearity dominance around the dip.
In these cases, $|\pi_-|\ll|\pi_+| e^{-3N_2}$ is achieved and thus the last term in parentheses in Eq.~\eqref{generalDelta} becomes large, which yields large $\Delta$.

\begin{figure} 
    \centering

    \includegraphics[width=
            0.9\hsize]{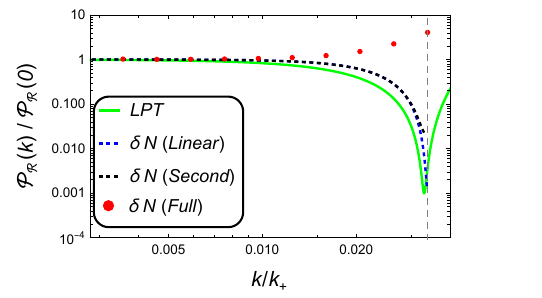}

    \includegraphics[width=
            0.9\hsize]{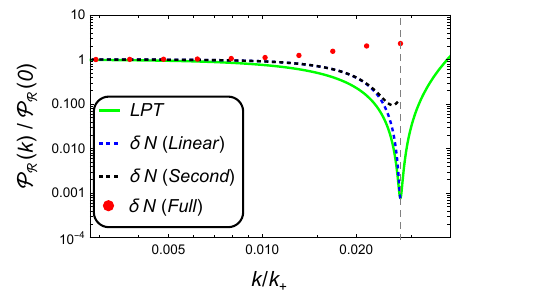} 
    \caption{\label{Fig_Stepmodel_Dip} The power spectrum of the curvature perturbation for the bump model (upper panel) and step model (lower panel) calculated by the linear perturbation theory (green solid line), and the variance of curvature perturbation calculated by the $\delta N$ formalism within linear order (blue dotted line), second order (black dotted line) and full order (red dots). 
    The gradient expansion parameter is set to $\sigma=\sigma_{\dip}=0.0339\text{ (bump), }0.0277\text{ (step)}$ (see Table~\ref{table3}).
    The vertical axis is normalised by the power spectrum at $k\to0$ and the horizontal axis is normalised by $k_+$.
    The vertical gray dashed line denotes the dip scale, which is the shortest scale we can calculate without considering the intrinsic non-Gaussianity in our model.
    }
\end{figure}

Finally, we comment on the validity of the perturbative expansion for the bump and step models.
In the above, we discussed the curvature perturbation up to the second order in $\delta X$ and $\delta Y$.
Let us compare the result from the second-order perturbation with that from the PDF, which includes full-order perturbations.
In Fig.~\ref{Fig_Stepmodel_Dip}, we plot the power spectrum and the variance of curvature perturbation for the bump and step models using the linear perturbation theory (green solid line) and the $\delta N$ formalism, respectively.
In the figure, we illustrate our calculations in the $\delta N$ formalism, truncated at the linear order (blue dotted line) and the second order (black dotted line), as well as the full order (red dots) calculated non-perturbatively by using the PDF.
The result from the non-perturbative calculation smoothly fills and erases the dip, and the variance at the dip scale is larger than the long-wavelength limit. On the other hand, the result from computation up to second order still has a shallow dip.
Thus, it is implied that, in models where the potential undergoes an abrupt change, as in the case of bump or step models, the perturbative analysis breaks down and a non-perturbative approach is required.
Note that our analysis does not include the $k^2$ correction and the Fourier mode mixing.
Therefore, we cannot rule out the possibility that the large non-linear effect we observed happened to be cancelled out by those effects we did not take into account, thereby restoring the validity of the perturbative expansion.
In order to more accurately estimate the statistical properties of the curvature perturbation, their effects also need to be included as in Refs.~\cite{Artigas:2024xhc,Briaud:2025ayt}, which is one of the remaining tasks.
In this paper, we adopted the $\delta N$ formalism as a non-perturbative method, but other methods are also available.
For example, non-perturbative lattice calculations in the context of inflation have begun to be studied in recent years (see, e.g., Refs.~\cite{Caravano:2021pgc,Caravano:2024tlp,Mizuguchi:2024kbl,Caravano:2024moy}).
We leave their application to general models, including bump and step models, for future work.

%

\section{Conclusion}
\label{conclusion}
We considered several characteristic models of single-field inflation in which the inflaton undergoes a non-slow-roll stage between two slow-roll stages.
The amplification of the curvature perturbation is known to occur, and a peak appears in such models. 
In many of those models, a dip also appears in the spectrum at a scale slightly larger than the peak scale.
In this paper, we focused on ``the dip'' and carefully studied its properties.
By using the $\delta N$ formalism extensively, we found that there is a linear combination of $\delta\varphi$ and $\delta\pi$, namely, $\delta X\,(=\delta\varphi$) and $\delta Y\,(=\delta\pi+g\delta\varphi$), whose cross-correlation vanishes.
In terms of these variables, we derived a perturbative formula~\eqref{deltaN2} for the curvature perturbation up through the second order.

In Sec.~\ref{linearanalysis}, the result from the $\delta N$ formalism is truncated at the linear level and compared with that of the standard linear perturbation theory.
This enabled us to interpret the condition for the appearance of a dip in the power spectrum in the $\delta N$ formalism language.
The dip condition~\eqref{kdipcondition} we found indicates that the expansion coefficient $N_X$ of the $\delta X$ perturbation vanishes at the dip. This implies that the contribution from the $\delta Y$ perturbation, which is subdominant on the other scales, makes the dominant contribution at the dip, hence must be taken into account.

In Sec.~\ref{nonlinearity}, we discussed the non-linearity of the curvature perturbation at the dip. Since the linear contribution of $\delta X$ disappears there, the higher-order contributions can be significantly important.
We introduced a parameter $\Delta$ that describes the non-linearity of the curvature perturbation with respect to its variance in Eq.~\eqref{Deltadef}, and evaluated its values in the models we considered.
The result shows that in the USR and Starobinsky's linear potential models, $\Delta$ is found to be small, which implies the dominance of the linear contribution from $\delta Y$, whereas in the bump and step models, $\Delta$ is larger than unity, which indicates the significance of the non-linear corrections. This was further confirmed by computing the \acp{PDF} numerically. 
The deviations from the Gaussian distribution are highly pronounced in the bump and step models.
As a result of the non-linear contributions, the dip that appeared in the linear power spectrum is found to disappear smoothly, as shown in Fig.~\ref{Fig_Stepmodel_Dip}.
As mentioned in the final paragraph of Sec.~\ref{nonlinearity}, we note that our analysis does not include the $k^2$ correction and the mode mixing, and therefore further analysis is required to draw firm conclusions.

Looking into the future, it will be important to see how the properties of the dip can be reflected in observable quantities.
On the theoretical side, since the bump and step models are found to have larger non-linearity than the USR and Starobinsky models, computing the quantum corrections and comparing them with the current $\delta N$ results may give us more insights into the non-linear effects in those models.
Related to this, it has been suggested that loop corrections from short wavelength modes to long wavelength modes are not large when the approximate consistency relations are well established~\cite{Tada:2023rgp,Kawaguchi:2024rsv,Fumagalli:2024jzz}.
Nevertheless, it is non-trivial whether the consistency relations are well established for long but finite wavelength modes when an abrupt change, such as a bump or step, exists in the inflaton potential.
We leave these as future works.

\acknowledgments
We thank Vadim Briaud and Vincent Vennin for useful discussions.
We also thank the Yukawa Institute for Theoretical Physics at Kyoto University, where the present work was started to be discussed during the long-term workshop ``Gravity and Cosmology 2024".
This work is supported by JSPS KAKENHI grants 
Nos.~23K03424 (TF), 
24KJ2108 (RK), 
20H05853 (MS), 24K00624 (MS), and 24K07047 (YT), and by the World Premier International Research Center Initiative (WPI Initiative), MEXT, Japan.


\renewcommand{\theequation}{A.\arabic{equation}}
\setcounter{equation}{0}

\appendix

\section{\boldmath Coefficients of $\delta N$}
\label{CoefficientsofdeltaN}
Here, we provide an explicit list of the coefficients of $\delta N$.
These results are easily obtained by perturbing the total 
\efolding number given in Eq.~\eqref{totalefolds} in terms of $\delta\varphi$ and $\delta\pi$.
Three patterns are possible depending on when the starting point, i.e., the scale of interest, leaves the Hubble horizon, and these are categorised and listed below.
\begin{itemize}
\item  Case for starting point in 1st stage ($k\leq \sigma k_+$)
\bae{
&
N_{\varphi}=\frac{1}{A_1}\,,
\label{N_varphi_first}
\\
&\begin{aligned}
N_{\pi}=\bmte{\frac{e^{-3N_1}}{3A_1 A_3(A_1+A_2(e^{3N_2}-1))} \\
\times\left(A_1^2+A_2A_3(e^{3N_1}-1)(e^{3N_2}-1)+A_1A_3(e^{3N_1}+e^{3N_2}-2)\right)\,,}
\label{N_pi_first}
\end{aligned}
\\
&
N_{\varphi\varphi}=0\,,
\\
&
N_{\varphi\pi}=\frac{e^{-3N_1}}{A_1^2 A_3 \left(A_2(1-e^{3N_2})-A_1\right)}\left(A_1^2-2A_1 A_3+A_2 A_3+(A_1-A_2)A_3 e^{3N_2}\right)\,,
\\
&\begin{aligned}
N_{\pi\pi}=\bmte{-\frac{e^{-6N_1}}{3A_2^2}\Biggl(\frac{2(A_1-A_2)A_2(e^{3N_1}-1)}{A_1^2}+\frac{(A_2-A_3)\left(A_1+A_2(2e^{3N_1}-1)\right)}{A_3\left(A_1+A_2(e^{3N_2}-1)\right)} 
\\
-\frac{2A_1^2(A_2-A_3)}{A_3\left(A_1+A_2(e^{3N_2}-1)\right)^2}+\frac{A_1^2(A_1-A_2)(A_2-A_3)}{A_3\left(A_1+A_2(e^{3N_2}-1)\right)^3}\Biggr)\,,}
\label{N_pipi_first}
\end{aligned}
}
where $N_1=\ln{(\sigma k_+/k)}$ and $N_2=\ln{(k_-/k_+)}$.

\item Case for starting point in 2nd stage ($\sigma k_+<k\leq\sigma k_-$)
\bae{
&
N_{\varphi}=\frac{A_1-A_2+A_3 e^{3N_2}}{A_3 \left(A_1+A_2(e^{3N_2}-1)\right)}\,,
\\
&
N_{\pi}=\frac{A_1+A_2(e^{3N_0}-1)+A_3 (e^{3N_2}-e^{3N_0})}{3A_3 \left(A_1+A_2(e^{3N_2}-1)\right)}\,,\\
&
N_{\varphi\varphi}=-\frac{3(A_1-A_2)(A_2-A_3)e^{6N_2}}{A_3\left(A_1+A_2(e^{3N_2}-1)\right)^3}\,,
\\
&
N_{\varphi\pi}=-\frac{(A_2-A_3)\left(A_1+A_2(e^{3N_0}-1)\right)e^{6N_2}}{A_3\left(A_1+A_2(e^{3N_2}-1)\right)^3}\,,
\\
&
N_{\pi\pi}=-\frac{(A_2-A_3)(e^{3N_2}-e^{3N_0})\left((A_1-A_2)e^{3N_0}+e^{3N_2}\left(A_1+A_2(2e^{3N_0}-1)\right)\right)}{3A_3\left(A_1+A_2(e^{3N_2}-1)\right)^3}\,,
}
where $N_0=\ln{(k/\sigma k_+)}$ and $N_2=\ln{(k_-/k_+)}$.
\item Case for starting point in 3rd stage ($k>\sigma k_-$)
\bae{
&
N_{\varphi}=\frac{1}{A_3}\,,
\\
&
N_{\pi}=\frac{1}{3A_3}\,,\\
&
N_{\varphi\varphi}=
N_{\varphi\pi}=
N_{\pi\pi}=0\,.
}
\end{itemize}

\section{Figures for Starobinsky, Bump, and Step models}
\label{figures_other_models}
\begin{figure} 
    \centering
    \includegraphics[width=0.8\hsize]{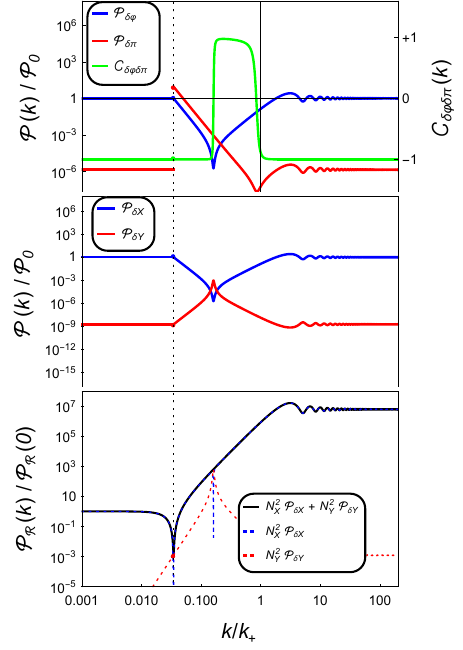}
    \caption{\label{Fig_powerspectra_Starobinsky} 
    \emph{Top panel}: the power spectra of $\delta\varphi$ and $\delta\pi$, and the correlation function of $\delta\varphi$ and $\delta\pi$, defined in Eq.~\eqref{eq: power and correlation}, for Starobinsky's linear potential model with model parameters shown in the first row of Table~\ref{table1}.
    They are evaluated when $k=\sigma aH$ with $\sigma=0.0347$. The two vertical dotted lines represent $k=\sigma k_+$ and $k=\sigma k_-$.
    The normalization factor ${\cal{P}}_0$ is $H^2/4\Pi^2 \Mpl^2$.
    \emph{Middle panel}: the power spectra of $\delta X$ and $\delta Y$.
    \emph{Bottom panel}: the power spectrum of curvature perturbation $\calR$ calculated by the $\delta N$ formalism at the linear order (black line).
    It is normalised by the power spectrum at $k\to0$.
    The components of Eq.~\eqref{PR}, $N_X^2\calP_{\delta X}$ and $N_Y^2\calP_{\delta Y}$, are also shown (blue and red dotted lines, respectively).
    Thanks to the choice $\sigma=\sigma_\dip=0.0347$ defined in Eq.~\eqref{sigma_dip}, the dip scale is located at $k=\sigma_\dip k_+$.
    }
\end{figure}

\begin{figure} 
    \centering
    \includegraphics[width=0.8\hsize]{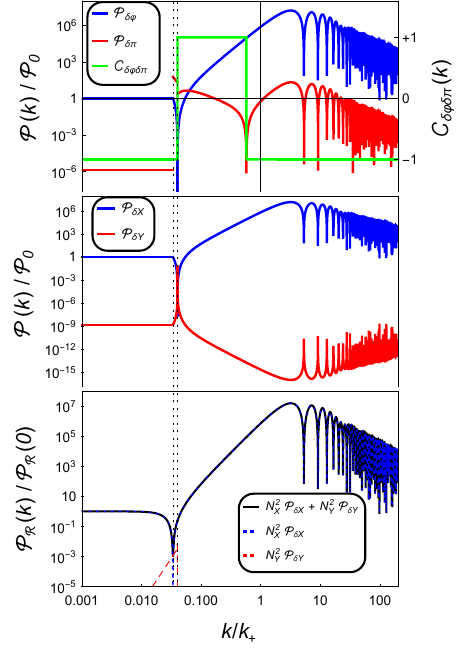}
    \caption{\label{Fig_powerspectra_Bump} 
    \emph{Top panel}: the power spectra of $\delta\varphi$ and $\delta\pi$, and the correlation function of $\delta\varphi$ and $\delta\pi$, defined in Eq.~\eqref{eq: power and correlation}, for the bump model with model parameters shown in the first row of Table~\ref{table1}.
    They are evaluated when $k=\sigma aH$ with $\sigma=0.0339$. The two vertical dotted lines represent $k=\sigma k_+$ and $k=\sigma k_-$.
    The normalization factor ${\cal{P}}_0$ is $H^2/4\Pi^2 \Mpl^2$.
    \emph{Middle panel}: the power spectra of $\delta X$ and $\delta Y$.
    \emph{Bottom panel}: the power spectrum of curvature perturbation $\calR$ calculated by the $\delta N$ formalism at the linear order (black line).
    It is normalised by the power spectrum at $k\to0$.
    The components of Eq.~\eqref{PR}, $N_X^2\calP_{\delta X}$ and $N_Y^2\calP_{\delta Y}$, are also shown (blue and red dotted lines, respectively).
    Thanks to the choice $\sigma=\sigma_\dip=0.0339$ defined in Eq.~\eqref{sigma_dip}, the dip scale is located at $k=\sigma_\dip k_+$.
    }
\end{figure}

\begin{figure} 
    \centering
    \includegraphics[width=0.8\hsize]{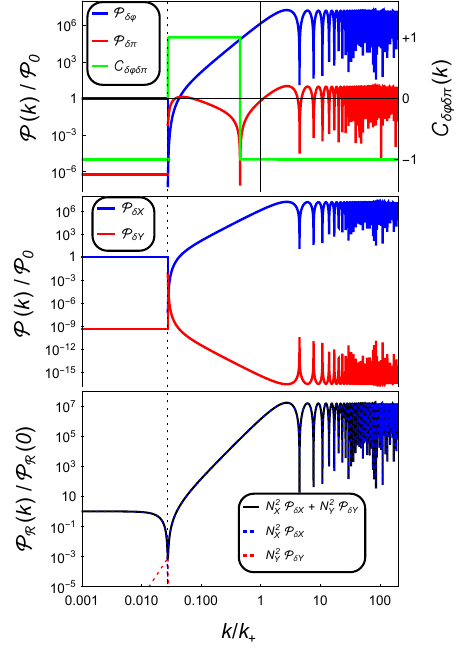}
    \caption{\label{Fig_powerspectra_step} 
    \emph{Top panel}: the power spectra of $\delta\varphi$ and $\delta\pi$, and the correlation function of $\delta\varphi$ and $\delta\pi$, defined in Eq.~\eqref{eq: power and correlation}, for the step model with model parameters shown in the first row of Table~\ref{table1}.
    They are evaluated when $k=\sigma aH$ with $\sigma=0.0277$. The two vertical dotted lines represent $k=\sigma k_+$ and $k=\sigma k_-$.
    The normalization factor ${\cal{P}}_0$ is $H^2/4\Pi^2 \Mpl^2$.
    \emph{Middle panel}: the power spectra of $\delta X$ and $\delta Y$.
    \emph{Bottom panel}: the power spectrum of curvature perturbation $\calR$ calculated by the $\delta N$ formalism at the linear order (black line).
    It is normalised by the power spectrum at $k\to0$.
    The components of Eq.~\eqref{PR}, $N_X^2\calP_{\delta X}$ and $N_Y^2\calP_{\delta Y}$, are also shown (blue and red dotted lines, respectively).
    Thanks to the choice $\sigma=\sigma_\dip=0.0277$ defined in Eq.~\eqref{sigma_dip}, the dip scale is located at $k=\sigma_\dip k_+$.
    }
\end{figure}

\begin{figure}
    \centering
    \begin{tabular}{c}
        \begin{minipage}{0.95\hsize}
            \centering
            \includegraphics[width=0.9\hsize]{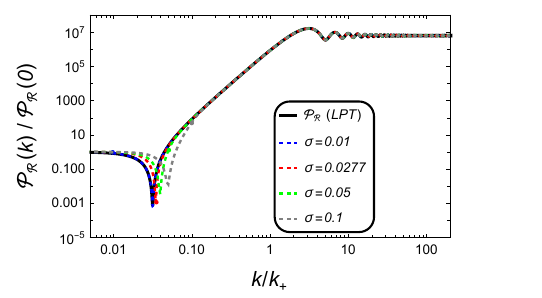}
        \end{minipage}
        \\
        \\
        \begin{minipage}{0.95\hsize}
            \centering
            \includegraphics[width=0.9\hsize]{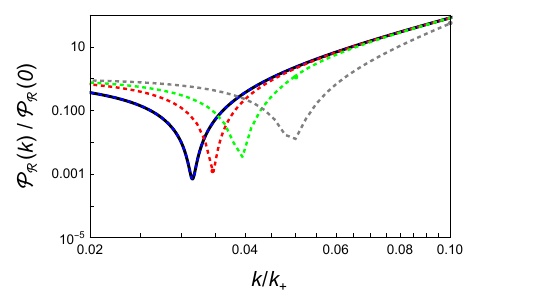}
        \end{minipage}
    \end{tabular}
    \caption{\label{powerspectrumsigma_Starobinsky}  
    \emph{Upper}: the power spectrum of curvature perturbation $\calR$ for the Starobinsky's linear potential model (see Table~\ref{table1} in Sec.~\ref{Appendixpowerspectrum} for detailed parameters) calculated by linear perturbation theory (black solid line) and $\delta N$ formalism with various gradient expansion parameter $\sigma$ (coloured dotted lines).
    These are normalised by the power spectrum at $k=0$.
    \emph{Lower}: enlarged version of the power spectrum around the dip scale.}
\end{figure}

\begin{figure}
    \centering
    \begin{tabular}{c}
        \begin{minipage}{0.95\hsize}
            \centering
            \includegraphics[width=0.9\hsize]{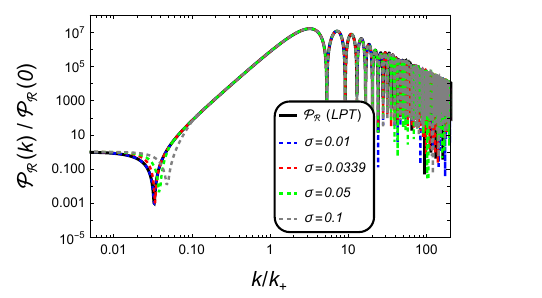}
        \end{minipage}
        \\
        \\
        \begin{minipage}{0.95\hsize}
            \centering
            \includegraphics[width=0.9\hsize]{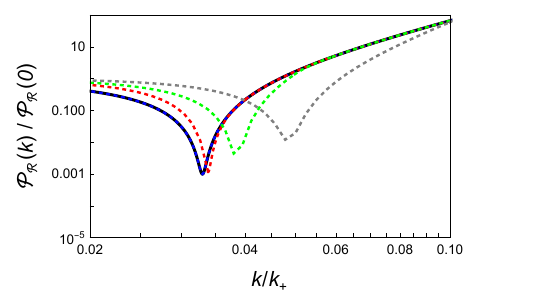}
        \end{minipage}
    \end{tabular}
    \caption{\label{powerspectrumsigma_bump}  
    \emph{Upper}: the power spectrum of curvature perturbation $\calR$ for the bump model (see Table~\ref{table1} in Sec.~\ref{Appendixpowerspectrum} for detailed parameters) calculated by linear perturbation theory (black solid line) and $\delta N$ formalism with various gradient expansion parameter $\sigma$ (coloured dotted lines).
    These are normalised by the power spectrum at $k=0$.
    \emph{Lower}: enlarged version of the power spectrum around the dip scale.}
\end{figure}

\begin{figure}
    \centering
    \begin{tabular}{c}
        \begin{minipage}{0.95\hsize}
            \centering
            \includegraphics[width=0.9\hsize]{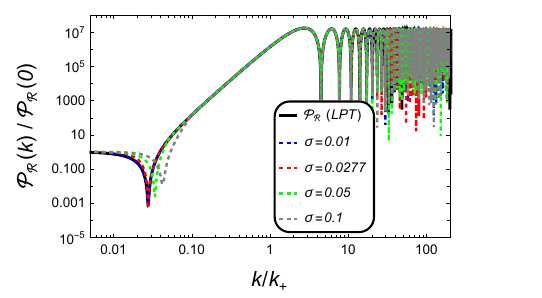}
        \end{minipage}
        \\
        \\
        \begin{minipage}{0.95\hsize}
            \centering
            \includegraphics[width=0.9\hsize]{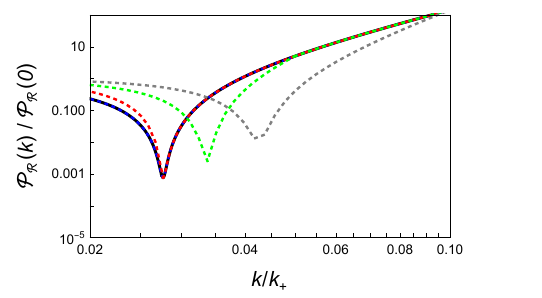}
        \end{minipage}
    \end{tabular}
    \caption{\label{powerspectrumsigma_step}  
    \emph{Upper}: the power spectrum of curvature perturbation $\calR$ for the step model (see Table~\ref{table1} in Sec.~\ref{Appendixpowerspectrum} for detailed parameters) calculated by linear perturbation theory (black solid line) and $\delta N$ formalism with various gradient expansion parameter $\sigma$ (coloured dotted lines).
    These are normalised by the power spectrum at $k=0$.
    \emph{Lower}: enlarged version of the power spectrum around the dip scale.}
\end{figure}

In this appendix, we show several figures for Starobinsky's linear potential, bump, and step models.
In Figs.~\ref{Fig_powerspectra_Starobinsky}--\ref{Fig_powerspectra_step}, the top panels show the power spectra of $\delta\varphi$ and $\delta\pi$ and the correlation function of $\delta\varphi$ and $\delta\pi$, the middle panels show the power spectra of $\delta X$ and $\delta Y$, and the lower panels show the power spectrum of the curvature perturbation calculated by $\delta N$ formalism.
The corresponding figures for the USR model can be found in Fig.~\ref{Fig_powerspectrum_XY}.
In Figs.~\ref{powerspectrumsigma_Starobinsky}--\ref{powerspectrumsigma_step}, we illustrate the power spectrum of the curvature perturbation obtained by the linear perturbation theory (black solid line) and the $\delta N$ formalism with various $\sigma$'s (coloured dotted lines).
The corresponding figures for the USR model can be found in Fig.~\ref{powerspectrumsigma}.


\clearpage

\bibliographystyle{JHEP}
\bibliography{bibJCAP}

\end{document}